\documentclass[prd,amsmath,amssymb,floatfix,nofootinbib,superscriptaddress]{revtex4}
\pdfoutput=1
\usepackage{graphicx}
\usepackage{bm}
\usepackage{amsmath,amssymb}
\usepackage{epsfig}
\usepackage{color}
\usepackage{natbib}
\usepackage{enumerate}
\usepackage{ifthen}

\usepackage[pdftex]{hyperref}

\def\eprinttmp@#1arXiv:#2 [#3]#4@{
\ifthenelse{\equal{#3}{x}}{\href{http://arxiv.org/abs/#1}{#1}}{\href{http://arxiv.org/abs/#2}{arXiv:#2} [#3]}}

\providecommand{\eprint}[1]{\eprinttmp@#1arXiv: [x]@}
\newcommand{\adsurl}[1]{\href{#1}{ADS}}
\providecommand{\bibinfo}[2]{\ifthenelse{\equal{#1}{isbn}}{
\href{http://cosmologist.info/ISBN/#2}{#2}}{#2}}

\newcommand{\tT}{\tilde{T}}
\newcommand{\Trispec}{{\cal T}}

\newcommand{\Mpc}{\text{Mpc}}
\newcommand{\half}{{\textstyle \frac{1}{2}}}

\newcommand{\fnl}{f_{\rm{NL}}}
\newcommand{\taunl}{\tau_{\rm{NL}}}
\newcommand{\gnl}{g_{\rm{NL}}}

\newcommand{\begm}{\begin{pmatrix}}
\newcommand{\enm}{\end{pmatrix}}

\newcommand\ba{\begin{eqnarray}}
\newcommand\ea{\end{eqnarray}}
\newcommand\bea{\begin{eqnarray}}
\newcommand\eea{\end{eqnarray}}

\newcommand\be{\begin{equation}}
\newcommand\ee{\end{equation}}

\newcommand{\la}{\langle}
\newcommand{\ra}{\rangle}

\newcommand{\ud}{{\rm d}}

\newcommand{\mC}{\bm{C}}

\newcommand{\boldvec}[1]{{{\mathbf{#1}}}}

\newcommand{\vK}{\boldvec{K}}

\newcommand{\vX}{\boldvec{X}}

\newcommand{\vc}{\boldvec{c}}

\newcommand{\vk}{\boldvec{k}}

\newcommand{\vx}{\boldvec{x}}

\newcommand{\cla}{\mathcal{A}}

\newcommand{\clo}{\mathcal{O}}

\begin{document}

%%%%%%%%%%%%%%%%%%%%%%%%%%%%%%%%%%%%%%%%%%%%%%%%%%%%%%%%%%%%%%%%%%
%                       Title matter                             %
%%%%%%%%%%%%%%%%%%%%%%%%%%%%%%%%%%%%%%%%%%%%%%%%%%%%%%%%%%%%%%%%%%
% title and affiliations

\title{The real shape of non-Gaussianities}

\author{Antony Lewis}
\address{Department of Physics \& Astronomy, University of Sussex, Brighton BN1 9QH, UK}
\homepage{http://cosmologist.info}

\begin{abstract}
I review what bispectra and trispectra look like in real space, in terms of the sign of particular shaped triangles and tetrahedrons. Having an equilateral density bispectrum of positive sign corresponds to having concentrated overdensities surrounded by larger weaker underdensities. In 3D these are concentrated density filaments, as expected in large-scale structure. As the shape changes from equilateral to flattened the concentrated overdensities flatten into lines (3D planes). I then focus on squeezed bispectra, which can be thought of as correlations of changes in small-scale power with large-scale fields, and discuss the general non-perturbative form of the squeezed bispectrum and its angular dependence. A general trispectrum has tetrahedral form and I show examples of what this can look like in real space. Squeezed trispectra are of particular interest and come in two forms, corresponding to large-scale variance of small-scale power, and correlated modulations of a bispectrum. Flattened trispectra can be produced by line-like features in 2D, for example from cosmic strings, and randomly located features also give a non-Gaussian signal. There are relationships between the squeezed types of non-Gaussianity, and also a useful interpretation in terms of statistical anisotropy. I discuss the various possible physical origins of cosmological non-Gaussianities, both in terms of primordial perturbations and late-time dynamical and geometric effects.
\end{abstract}

\date{\today}

\maketitle

\pagenumbering{arabic}

\section{Introduction}

Most of the fields that we observe in cosmology are significantly non-Gaussian: the large-scale matter density is strongly non-Gaussian due to non-linear growth, and even the CMB is significantly non-Gaussian on small scales due to non-linear effects. Current evidence suggests that primordial fluctuations at the beginning of the hot big bang were very close to Gaussian, but primordial non-Gaussianity is still a possibility and, if observed, would be a way to rule out wide classes of early-universe models. Understanding non-Gaussianity is therefore crucial to extract the most information from cosmological observations. Since the initial fluctuations were close to Gaussian, the non-Gaussianities can often be treated perturbatively, in which case the bispectrum and trispectrum contain most of the additional information (though at very late times where strongly non-linear processes are involved a deeper analysis may be required). Here I will focus on non-Gaussianity as described by a bispectrum and trispectrum. I will show what fields with various qualitatively different types of non-Gaussianity look like in real space, and give some general results for the form of the squeezed bispectrum, its angular decomposition, relation to the trispectrum, and the relation between anisotropy estimators and local non-Gaussianity.

This paper does not aim to review the details of observational analysis or physical modelling of non-Gaussianities; there are many excellent reviews and references available for further technical and physical model details (for a small sample see e.g. Refs. ~\cite{Bartolo:2004if,Creminelli:2006gc,Smith:2006ud,Yadav:2007rk,Fergusson:2008ra,Senatore:2009gt,Chen:2010xka,Byrnes:2010em,Komatsu:2010hc,Desjacques:2010nn}). I will mostly focus on generalities rather than specific models, though I mention specific cases of cosmological interest as we go along. For simplicity I restrict to scalar fields in flat space; the generalization to 2D fields on the sphere (e.g. the full-sky CMB), a non-flat background, or tensor fields (e.g. galaxy lensing and CMB polarization) is conceptually simple, though can be technically significantly more complicated.

I will assume a standard statistically isotropic and homogeneous background cosmology. In this case the wavevectors describing a particular non-Gaussian configuration of modes must sum to zero: for a bispectrum there are three modes with wavevectors that close to form a triangle; for a trispectrum there are four wavevectors that define four edges of a tetrahedron. By ``shape", in this paper I will mean a particular configuration of fixed-length wavevectors, e.g. a specific bispectrum triangle. A general bispectrum or trispectrum gives the full wavenumber and configuration dependence, i.e. the amount of signal expected in each possible shape. For example an `orthogonal' bispectrum does not correspond to a particular triangle shape, but instead a particular wavenumber dependence of the signal in each shape. The more general wavevector dependence of the bispectrum and trispectrum has been discussed extensively elsewhere~\cite{Fergusson:2008ra,Senatore:2009gt,Regan:2010cn}; instead of showing the magnitude of the signal as a function of the wavevectors, I instead focus on what the signal looks like in real space for specific configurations. This can be a useful aid to understanding which kinds of physical process generate the different shapes.

\section{Gaussianity and the power spectrum}

Before discussing non-Gaussianity, it is worth quickly remembering the key features of Gaussian fields. In particular we are usually interested is statistically isotropic and homogenous universe models, and hence in fields that have these symmetries. For simplicity I shall focus mainly on scalar fields 2D flat space, for example a slice through the matter density field or a small patch of the CMB, but almost everything generalizes to other cases such as full-sky observations. Assuming we can measure a field $T(\vx)$ as a function of position, in flat space these can be Fourier transformed and written as
\be
T(\vx) = \frac{1}{(2\pi)^{N/2}} \int \ud \vk\, T(\vk) e^{i\vk\cdot\vx}
\ee
where $N$ is the number of dimensions. Statistical homogeneity and isotropy means that the statistical properties of the field must be unchanged under translations and rotations $T(\vx)\rightarrow T(\vx')$, so $\la T(\vx) T(\vx')\ra$ can only be a function of the invariant separation between the points $|\vx-\vx'|$. This implies that the covariance of the field is determined by a power spectrum depending only on $k\equiv |\vk|$:
\be
\la T(\vk_1) T(\vk_2)\ra = \delta(\vk_1+\vk_2) P(k_1).
\ee
For a small patch of the CMB, the power spectrum is just $C_l$, where $l=k$. The delta function says that modes with different wavevectors are completely uncorrelated: knowing the sign of $T(\vk_1)$ tells you nothing about the likely sign of $T(\vk_2)$. From the power spectrum the only thing we know is the variance of each individual mode, which from the assumption of isotropy is the same independent of the orientation of the mode.

A purely Gaussian statistically homogeneous and isotropic field is fully described statistically by its power spectrum. However more interesting fields are possible that are also statistically homogeneous and isotropic, with the non-Gaussian statistics described by a series of higher-point correlation functions.

\section{Bispectrum}
The first non-Gaussian signal to consider is a bispectrum, corresponding to a three-point correlation, or in Fourier space a correlation between three different mode wavevectors. We are still interested in statistically homogeneous and isotropic fields, which implies the statistics are described by a reduced bispectrum $b(k_1,k_2,k_3)$ that depends only the lengths of the wavevectors:
\be
\la T(\vk_1) T(\vk_2)T(\vk_3)\ra = \frac{1}{(2\pi)^{N/2}} \delta(\vk_1+\vk_2+\vk_3) b(k_1,k_2,k_3).
\ee
From now on when referring to the bispectrum I will mean the reduced bispectrum as defined here; an analogous definition applies on the full sky. The delta-function here means that the 3-mode correlation is zero unless the wavevectors sum to zero: they form a triangle. If there is a non-zero bispectrum, modes with different wavevectors are not independent: if we measure $T(\vk_1)$ and $T(\vk_2)$, the sign of the bispectrum $b(k_1,k_2,k_3)$ then tells us which sign of $T(\vk_3)$ is more likely. Positive sign gives positive skewness (tail of very high values), negative sign gives negative skewness (tail of very low values). What this looks like in real space depends on the shape of the triangle (the relative lengths of the different wavevectors).

\subsection{Equilateral and flattened (folded) triangles}

\begin{figure}
\includegraphics[width=14cm]{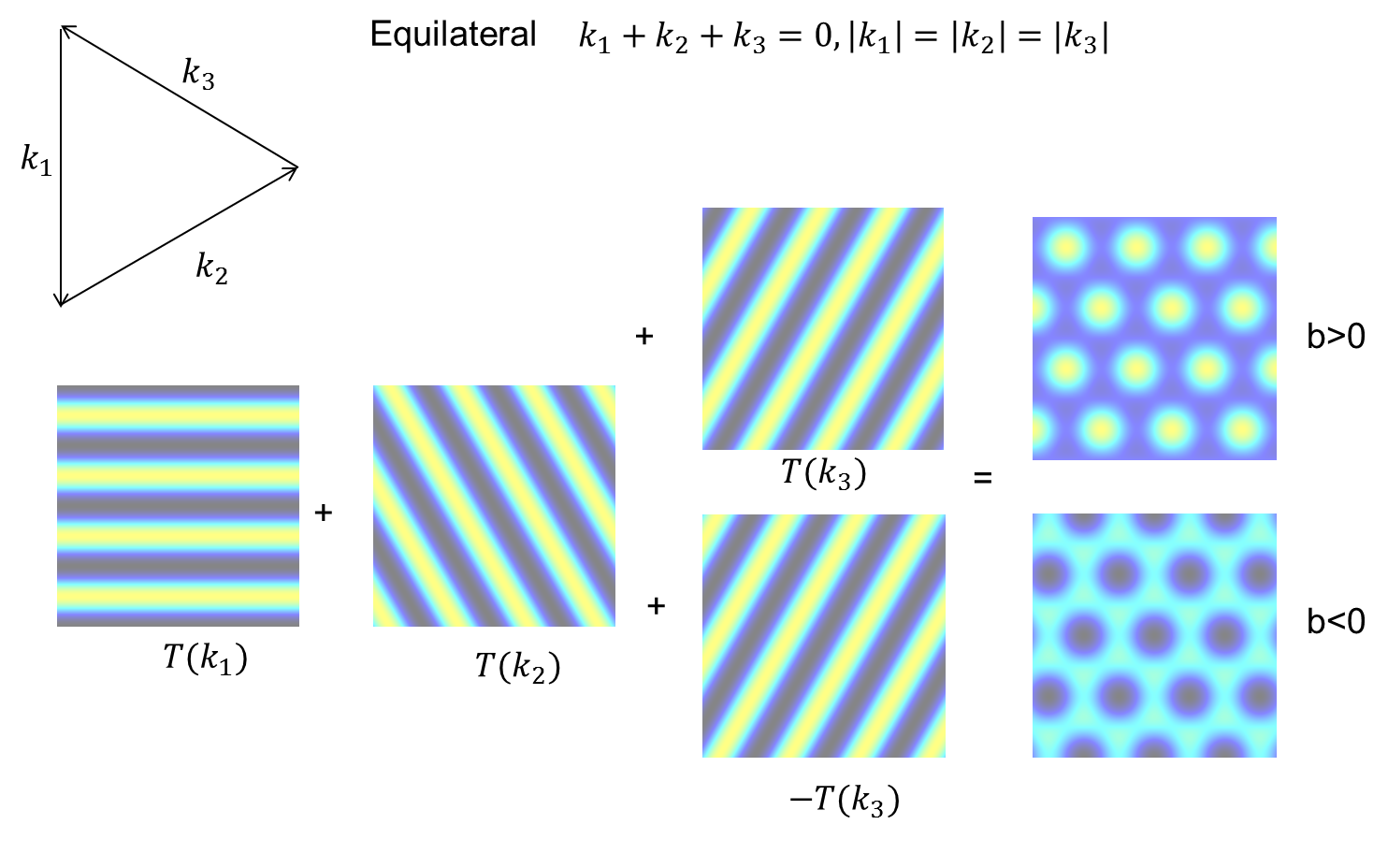}
\caption{Equilateral bispectrum: a field can be decomposed into plane-wave modes, and the three components with wavevectors that form an equilateral triangle may have different relative signs. The sign of the bispectrum tells you which combination of signs is more likely (on average gives a positive or negative product of the three modes). A positive reduced bispectrum corresponds to being likely to have waves combining to have strong overdensities surrounded by larger areas of milder underdensity. A negative equilateral bispectrum corresponds to being likely to have concentrated underdensities surrounded by areas of milder overdensity. Note that in 3D the figures extend into the page, and hence the positive bispectrum corresponds to concentrated overdense filaments surrounded by larger areas of milder underdensity.
\label{Equilateral}
}
\end{figure}

The first case I consider is equilateral triangles, where the lengths of the three sides of the triangle are the same, $k_1=k_2=k_3$. If there's a non-zero equilateral bispectrum, what does the field look like in real space? To answer this we can consider taking the $T(\vk_1)$ and $T(\vk_2)$ components of the field, and then ask what the bispectrum tells us about $T(\vk_3)$. Depending on the relative sign of $T(\vk_3)$, a field consisting of these three modes looks rather different --- see Fig.~\ref{Equilateral}. The sign of the bispectrum tells us which sign of $T(\vk_3)$ is more likely, in other words whether we are more likely to have small regions of concentrated overdensity ($b>0$) or regions of concentrated underdensity ($b<0$). As can easily be imagined, such patterns can be obtained by locally moving matter around, for example concentrated overdensities can form by gravitational collapse, and thus equilateral non-Gaussianity is likely to be present in any field undergoing local non-linear dynamical processes.

\begin{figure}
\includegraphics[width=10cm]{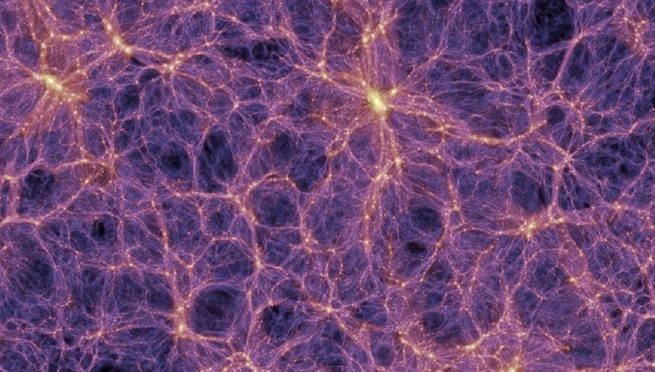}
\caption{A snapshot of non-linear large-scale structure from the millennium simulations~\cite{Springel:2005nw}.
Dynamical non-linear collapse of very dense filaments (surrounded by milder underdensities, voids) generates a large positive roughly equilateral density bispectrum.
%(Note that of course a bispectrum and power spectrum alone does not fully describe the statistics of this field; a bispectrum triangle defines a plane, and higher-order moments are therefore required to fully describe the statistics of objects such as galaxy clusters which are localized in 3D.)
\label{Millennium}
}
\end{figure}

A bispectrum is determined by three wavevectors which always lie in a plane. In 3D, the modes we are considering correspond to plane waves, and the concentrated overdensities correspond to filaments. These are precisely what form during the growth of large-scale structure, as shown in the famous simulation of Fig.~\ref{Millennium}. Since it is the overdensities that are concentrated, not the underdensities, the non-linear large-scale structure density field will have a large positive equilateral component to its bispectrum (for a perturbation theory calculation see Ref.~\cite{Fry:1983cj}).

\begin{figure}
\includegraphics[width=10cm]{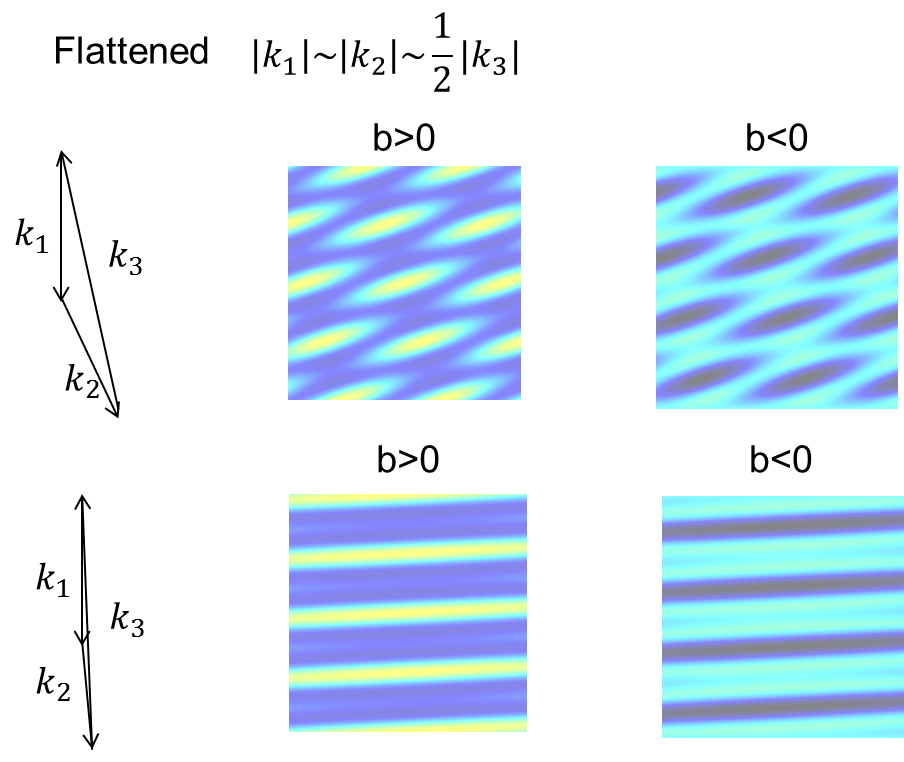}
\caption{As an approximately equilateral bispectrum triangle flattens, the round areas of overdensity become flattened into pancakes. In 3D a positive flattened bispectrum with $k_1=k_2=k_3/2$ corresponds to being likely to have overdense pancakes with larger mildly underdense planes in between.
\label{Flattened}
}
\end{figure}

Of course exactly equilateral triangles are a very special case, but there are many shapes that are close to equilateral and these will also look similar, but correspond to slightly elliptical concentrated overdensities or underdensities. As the bispectrum triangle becomes more flattened, these turn into a line, or in 3D concentrated overdensity or underdensity pancakes (planes); see Fig.~\ref{Flattened}. Note that shapes that are qualitatively distinct in 3D may not be after projection into 2D: for example if an purely equilateral shape is present in 3D, projecting down to 2D will give flattened contributions when the line of sight lies close to the plane of the triangle (slicing a 3D filament along its length gives a line of overdensity).

Since equilateral-form non-Gaussianity involves wavevectors of roughly the same magnitude, these modes would have left the horizon during inflation at roughly the same time. Non-linear dynamics prior to horizon exit during inflation can therefore generate a bispectrum with a significant equilateral bispectrum component. By an analogy with large-scale structure growth, one might imagine than any non-linear growth prior to horizon exit would require a low sound speed for the perturbations\footnote{The sound speed, given by $c_s^2\equiv \delta p/\delta\rho$, measures the pressure perturbation $\delta p$ induced by a given density perturbation $\delta\rho$. In structure growth pressures prevent gravitational collapse. The sound horizon roughly determines the Jeans scale below which pressure prevents collapse, so a low sound speed is required in order to have much growth on small sub-horizon scales.}; however in standard single-field inflation the rest-frame sound speed of the perturbations is exactly the speed of light. This prevents significant non-Gaussianity developing. However in extended models the speed of sound can be much lower, and in such cases significant equilateral non-Gaussianity can develop~\cite{Senatore:2009gt}. Non-linear dynamics after horizon re-entry will of course also generate equilateral non-Gaussianity, for example second-order effects prior to recombination. Though small compared to the strongly non-linear growth of structure at lower redshifts, these signals might present an important source of possible confusion for small inflationary signals~\cite{Pitrou:2010sn}.

\subsection{Squeezed triangles}

Squeezed triangles correspond to having one wavevector much shorter than the other two: in other words one large-scale mode and two much shorter-scale modes.  The bispectrum is invariant under permutations of $k_1,k_2,k_3$, so for squeezed triangles it is convenient to adopt the convention that we permute indices so that $k_1\le k_2\le k_3$, and $k_1$ therefore always labels the large-scale mode.
 Sometimes people refer to the ``squeezed-limit", meaning the limit as $k_1\rightarrow 0$, but this is not really observationally relevant as very super-horizon modes are unobservable. By squeezed I will mean triangles with $k_2,k_3\gg k_1$, but the wavelength of $k_1$ not much larger than the horizon size today, so that the mode is still observationally relevant.

\begin{figure}
\includegraphics[width=14cm]{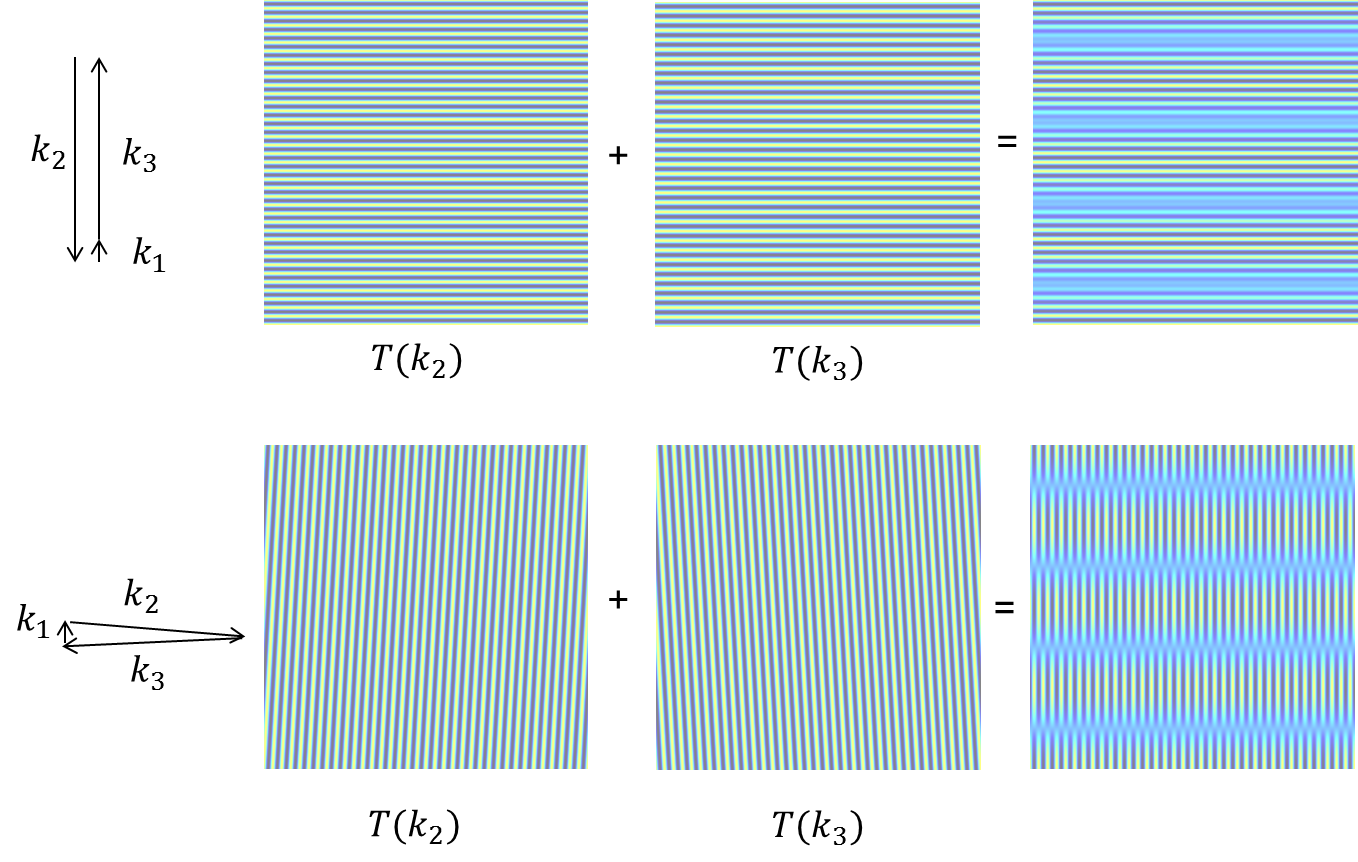}
\caption{Two short-scale modes combine giving interference patterns. These look like a large-scale modulation in the amplitude of the small-scale modes, with the modulation having wavevector $\vk_1=-\vk_2-\vk_3$.
\label{TwoWaves}
}
\end{figure}

\begin{figure}
\includegraphics[width=14.7cm]{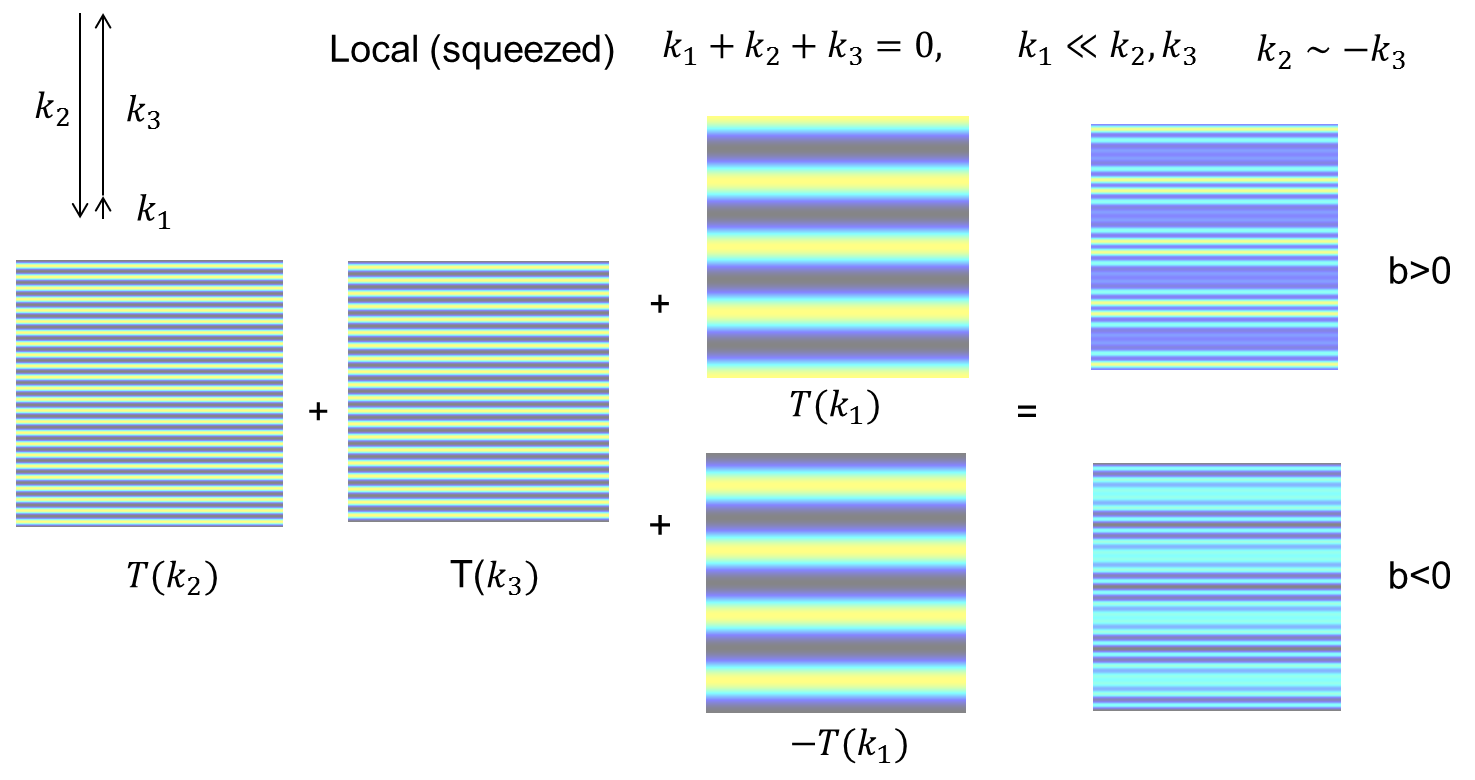}
\includegraphics[width=14cm]{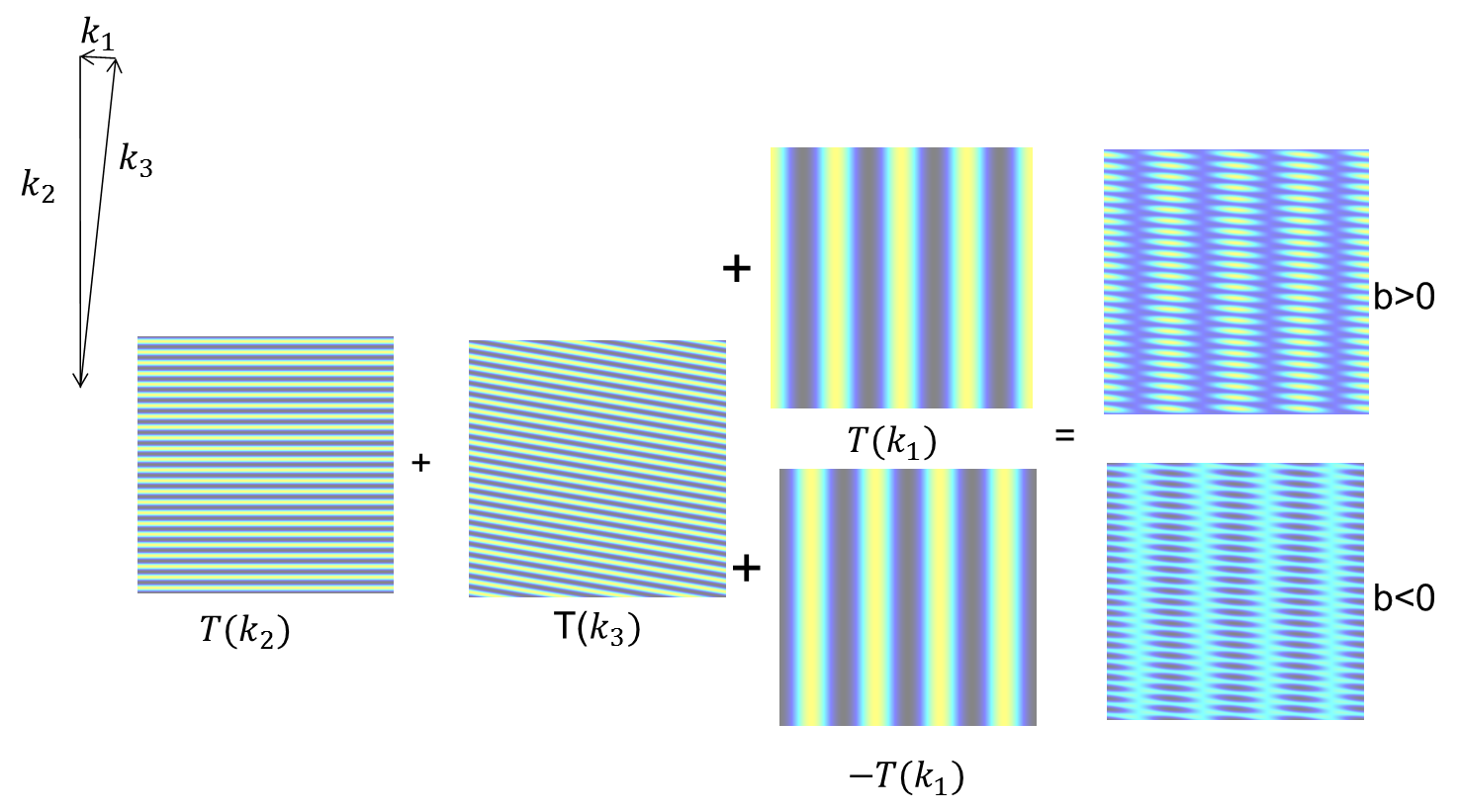}
\caption{Squeezed bispectrum: two small-scale modes with nearly-equal wavelength ($k_2\sim k_3$) interfere with each other, giving some regions with lots of small-scale power and others with destructive interference giving little small-scale power. The sign of the bispectrum tells you whether a region of high small-scale power is more likely to be associated with a large-scale overdensity or a large-scale underdensity. If the correlation is independent of the relative orientation (upper and lower figures have the same signal) the bispectrum is isotropic, but in general it is not.
\label{SqueezedAligned}
}
\end{figure}

First it is helpful to consider what a combination of two small-scale modes with $\vk_2\sim -\vk_3$ looks like: as shown in Fig.~\ref{TwoWaves} the waves destructively interfere in some regions leaving little small-scale structure, but in other regions they reinforce each other giving a large small-scale signal. So this looks like a large-scale modulation in the small-scale power, where the wavevector of the modulation is given by $-(\vk_2+\vk_3)$. In a Gaussian field the signs of all the modes are independent, so of course there is no modulation on average. However if there is a non-zero squeezed bispectrum, there is a correlation between this modulation and the large-scale modes; for example see Fig.~\ref{SqueezedAligned}.
 A positive squeezed bispectrum means that where there's a large-scale overdensity there's likely to be more small-scale structure, and where there is an large-scale underdensity there is likely to be less small-scale structure.

\subsubsection{Angular dependence and the squeezed approximation}

\begin{figure}
\includegraphics[width=14cm]{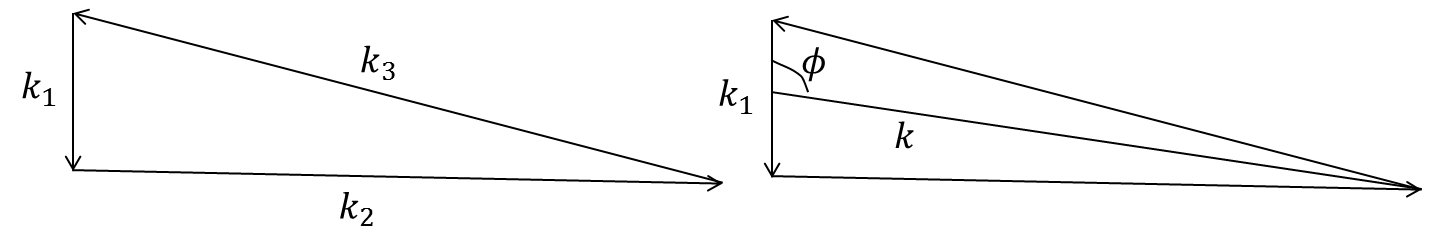}
\caption{A bispectrum triangle can be described by the lengths of the three sides, or alternatively by the length of the shortest side (corresponding to the large-scale mode), the length-scale of the long sides (short-scale modes) defined by  $k=|\vk_2-\vk_3|/2$, and the angle $\phi$ that measures the relative orientation of the long and short-scale modes.
\label{Triangles}
}
\end{figure}

So far we have described the bispectrum triangles in terms of the lengths of the sides. However for squeezed triangles in particular it can be useful to describe the triangles in a different way. Since the two small-scale modes are of similar wavelength, is natural to use a single number $k\equiv|\vk_2-\vk_3|/2$ to quantify the scale of the small-scale modes~\cite{Lewis:2011fk,Creminelli:2011rh}, as shown in Fig.~\ref{Triangles}. The remaining free parameter can then be taken to be the angle $\phi$ between the large-scale and small-scale wavevectors, so that a triangle is fully described by the three numbers $k_1, k, \phi$. Squeezed triangles have $k_1\ll k$, and in particular I shall refer to the squeezed limit as having leading corrections of ${\cal O}((k_1/k)^2)$. Note that exchanging $k_2\leftrightarrow k_3$ is equivalent to $\phi \leftrightarrow \pi-\phi$ with the same $k$.

It can also be useful to decompose the bispectrum depending on its angular dependence, i.e. writing\footnote{For bispectra in 3D, one could also expand into more directly orthogonal spherical harmonic $Y_{L0}$ modes; the argument is much the same so for simplicity I will stick with the 2D modes.}
\be
 b(k_1,k,\phi) = \sum_m b_{k_1,k}^{m} \, e^{ i m \phi}.
\ee
If the squeezed bispectrum is independent of $\phi$, i.e. the orientation between the large and small-scale modes, it is called isotropic. In this case the bispectrum is fully determined by the $m=0$ component $b_{k_1,k}^{0}$. For 3D or statistically parity-invariant 2D fields than can be no $\sin(m\phi)$ dependence, and
from rotational invariance the odd $m$ components should vanish, but in general there can be an angular dependence  $\cos(m\phi)$ with even $m$. This generally enters from
$\vk_1\cdot \vk_2$ and $\vk_1\cdot \vk_3$ dependence of the bispectrum, for example from gradient contractions,
and hence angular dependence enters in the combination $k_1\cos(\phi)$.
%If we expand $k_2$ and $k_3$ in powers of $k_1/k$, using $\vk\equiv(\vk_2-\vk_3)/2=\vk_2+\vk_1/2=-\vk_3-\vk_1/2$ we have
%\ba
%k_2 &=&k\sqrt{1 +(k_1/k)^2/4 - (k_1/k) \cos(\phi) }
%= k\left(1 - \frac{\cos(\phi)}{2} \frac{k_1}{k} + \frac{1-\cos(2\phi)}{16}\frac{k_1^2}{k^2}+\dots\right)
%\\
%k_3 &=&k\sqrt{1 +(k_1/k)^2/4 + (k_1/k) \cos(\phi) },
%= k\left(1 + \frac{\cos(\phi)}{2} \frac{k_1}{k} + \frac{1-\cos(2\phi)}{p16}\frac{k_1^2}{k^2}+\dots\right).
%\ea
%  and
A bispectrum expansion in $k_1/k$ therefore has angular dependence entering via powers of the dimensionless parameter $\epsilon_\phi\equiv (k_1/k)\cos(\phi)$. Since any bispectrum must be symmetric under $k_2\leftrightarrow k_3$ (and therefore $\phi\leftrightarrow \pi-\phi$), only even powers of $\epsilon_\phi$ can enter, and higher even powers of $\cos(\phi)$ (and hence higher even $m$) are suppressed by proportionally more powers of $k_1/k$.
 Thus the squeezed expansion of a smooth scalar bispectrum is typically of the form
\be
b(k_1,k_2,k_3)  = A(k_1,k)  + \left[B(k_1,k)+C(k_1,k)\cos(2\phi)\right] \frac{k_1^2}{k^2} + D(k_1,k,\phi)\clo(k_1^4/k^4),
\ee
where $A, B,C,D$ encode the scale-dependence of the particular physics involved. Hence unless there is very strong
scale dependence the leading term is isotropic, and the next leading term only has isotropic and quadrupolar contributions. For some bispectra $A=0$ and the leading term then in general has both isotropic and quadrupolar components. Physically the reason higher angular dependence does not appear is because a small patch on a large-scale scalar modulating field will be accurately described by a field value and gradient, with the field value giving an isotropic change to the small-scale power, and the gradient defining the local basis for the quadrupolar dependence of the small-scale power. The leading term can only have anisotropic contributions if the modulation is not scalar.

%%%%If a constant change in field value is not physical, then $A$ can be defined to be zero; for example a uniform lensing
%potential gives zero magnification and shear, or a uniform

For example gravitational lensing of the CMB generates both $m=0$ part of the bispectrum (corresponding to large-scale lenses isotropically magnifying and de-magnifying the CMB), and also an $m=2$ component from lensing shear (see e.g. Refs.~\cite{Creminelli:2004pv,Boubekeur:2009uk,Bucher:2010iv,Lewis:2011fk}). Primordial bispectra involving gravitational waves can also generate small $m=2$ components, since the modulation of the small-scale modes will depend on their orientation with respect to the anisotropic distortion produced by a large-scale tensor mode~\cite{Maldacena:2002vr}. Similarly inflation models with vector fields can generate anisotropic bispectra (e.g.~\cite{Barnaby:2012tk}). Purely scalar local modulations are expected to give isotropic bispectra, as we discuss further below, though leading terms from small gradients will give quadrupolar contributions (see e.g. Ref.~\cite{Burrage:2011hd}).

The decomposition of the squeezed bispectrum into angular moments is conceptually useful for distinguishing different physical effects. In particular, in the squeezed limit bispectra with different $m$ are orthogonal\footnote{In 2D, see Ref.~\cite{Pearson:2012ba}. For bispectra in 3D, one can expand in spherical harmonics $Y_{L0}$, and the modes with different $L$ will then be orthogonal.}: an estimator for $b_{k_1,k}^{m}$ should not be liable to confusion with a bispectrum $b_{k_1,k}^{m'}$ if $m\ne m'$. For example this is partly what allows CMB lensing to be easily distinguished from an isotropic local primordial bispectrum: the $m=0$ part due to magnification is a source of confusion, but the $m=2$ lensing signal is distinctive and allows the lensing to be isolated and subtracted. The angular decomposition is also useful when considering secondary processing of primordial bispectra: if a statistically isotropic small-scale process affects the primordial modes it will not mix bispectra of different $m$. For example under gravitational lensing any primordial squeezed $b_{k_1,k}^{m}$ is blurred out in $k$ due to random small-scale lensing deflections, but the angular $m$ dependence does not change~\cite{Hanson:2009kg,Pearson:2012ba}. Since all bispectra are expected only to have the lowest $m$ components in the squeezed limit, this also suggests that a general modal bispectrum decomposition can efficiently capture the angular dependence with only two angular modes for squeezed shapes, and since corrections are quadratic triangles do not need to be very squeezed for the squeezed approximation to be quite accurate.

Models in which statistical isotropy is broken can also generate anisotropic bispectra in the more general sense that the bispectrum signal can then depend on the orientation of the triangle\cite{Karciauskas:2008bc,Dimastrogiovanni:2010sm}; I will not consider this possibility further here.

\subsubsection{General form of the squeezed bispectrum}

Consider the case where the field being observed $\tilde{T}$ can be calculated from some non-linear function of a set of purely Gaussian fields $\{ X_i\}$ that we can write as a vector $\vX$. For example the Gaussian fields might be combinations of small scale and large-scale perturbation modes, different inflation field perturbations, lensing potentials, linear matter densities, velocity fields giving redshift distortions, etc. Since we are assuming underlying Gaussian fields, they are fully described by their covariance $\mC$ (a matrix of power spectra). If we are interested in the bispectrum of a field $\tilde{T}(X)$ (where the tilde denotes that it is non-linear), it can therefore be calculated by integrating out the Gaussian fields
\be
\la \tT(\vk_1) \tT(\vk_2) \tT(\vk_3) \ra = \int \ud \vX \frac{\exp\left(- \half \vX^\dag \mC^{-1} \vX \right) }{|2\pi \mC|^{1/2}}\tT(\vk_1) \tT(\vk_2) \tT(\vk_3).
\label{gaussint}
\ee
For a squeezed bispectrum the large-scale field $T(\vk_1)$ is at a much larger scale than the two small-scale fields $T(\vk_2)$, $T(\vk_3)$, and hence is often well approximated as being Gaussian (a linear function of the underlying Gaussian fields). For example in large-scale structure the large-scale mode in the bispectrum will be nearly linear if $k_1 \alt 0.05\Mpc^{-1}$, similarly in the CMB large scales are accurately linear but smaller scales have significant non-linearities due to lensing and other effects. The approximation $\tilde{T}(\vk_1)\approx T(\vk_1)$ is the linear short-leg approximation~\cite{Lewis:2011fk}, which is very accurate in some cases.
Writing $T(\vk_1)$ as a linear combination of $X_i(\vk_1)$ in Eq.~\eqref{gaussint}, $T(\vk_1)=M_i X_i(\vk_1)$, and then writing $X_i(\vk_1)$
as a functional derivative of the exponent,
\ba
\la T(\vk_1) \tT(\vk_2) \tT(\vk_3) \ra &=& \int \ud \vX \frac{\exp\left(- \half \vX^\dag \mC^{-1} \vX \right) }{|2\pi \mC|^{1/2}} M_i X_i(\vk_1) \tT(\vk_2) \tT(\vk_3) \nonumber\\
&=& -\int \ud \vX    M_i  C_{ij}(k_1)\frac{\delta}{\delta X_j(\vk_1)^*}\left( \frac{\exp\left(- \half \vX^\dag \mC^{-1} \vX \right) }{|2\pi \mC|^{1/2}}\right)\tT(\vk_2) \tT(\vk_3)\nonumber\\
&=&  M_i  C_{ij}(k_1) \int \ud \vX   \frac{\exp\left(- \half \vX^\dag \mC^{-1} \vX \right) }{|2\pi \mC|^{1/2}}\frac{\delta}{\delta X_j(\vk_1)^*}\left( \tT(\vk_2) \tT(\vk_3)\right).
\ea
Then since $C_{ij}  = P_{X_iX_j}$, we have $M_i  C_{ij} =P_{T X_j}$ and hence
\be
\la \tT(\vk_1) \tT(\vk_2) \tT(\vk_3) \ra \approx \la T(\vk_1) \tT(\vk_2) \tT(\vk_3) \ra =  P_{TX_i}(k_1) \left\la \frac{\delta}{\delta X_i(\vk_1)^*} \left(\tT(\vk_2) \tT(\vk_3)\right) \right\ra.
\label{squeezed_nonpert}
\ee
Thus the bispectrum depends on the correlation of the large-scale field with the modulating fields $P_{TX_i}(k_1)$, and is proportional to the response of the small-scale \emph{non-linear} modes to changes in the large-scale modulation. The result of Eq.~\eqref{squeezed_nonpert} is fully non-perturbative and only relies on the linear short-leg approximation, not an extreme squeezed limit (if necessary it can be generalized out of the linear short-leg approximation by including higher derivative terms). In the case of CMB lensing it is possible to calculate the response term essentially exactly non-perturbatively~\cite{Lewis:2011fk}; more generally as $k_1/k\rightarrow 0$, $k_2\rightarrow k_3$, and the response term just describes how the small-scale power spectrum changes with a different large-scale background modulation.

\subsubsection{Local scalar modulations}

Perhaps the simplest way to generate a bispectrum is from a purely local modulation of a Gaussian field.
 For example
\be
\chi(\vx) = \chi_0(\vx)[1+\psi(\vx)]
\label{modulation}
\ee
where both $\chi_0$ and $\psi$ are Gaussian fields. I will assume that $|\psi|\ll 1$ so that the modulation is small, consistent in the primordial context with the relatively tight observational bounds on non-Gaussianity of the CMB. To have a non-zero bispectrum the fields must be correlated, so $P_{\chi_0\psi}\ne 0$. The correlated part of $\psi$ is then conventionally written as $\frac{3}{5} \fnl \chi_0$, where $\fnl$ is an amplitude determined by the degree of correlation (and hence governing the size of the bispectrum). It is then straightforward to calculate the bispectrum to leading order in the modulation, giving
\be
b(k_1,k_2,k_3) = 2\frac{3}{5}\fnl[P(k_1)P(k_2)+P(k_1)P(k_3) +  P(k_2)P(k_3)]
\ee
where $P(k)$ is the power spectrum of $\chi_0$.

For primordial perturbations usually nearly scale-invariant fields are of interest, what does that mean in terms of $k$-dependence of $P(k)$? The total variance at any point in an N-dimensional field is  $\propto \int \ud^N k P(k) \propto \int  k^N P(k) \ud \ln k$. A scale-invariant spectrum has equal contributions to the variance from each dimensionless $\ud \ln k$ interval and hence has $P(k) \propto k^{-N}$. If the fields are close to scale invariant, for squeezed triangles with $k_2,k_3\gg k_1$ it follows that $P(k_2)P(k_3) \ll P(k_1)P(k_2) \approx P(k_1)P(k_3)$. In fact doing an expansion in $k_1/k$ for a scale-invariant spectrum we have
\bea
b(k_1,k_2,k_3) &\approx& 2\frac{3}{5}\fnl P(k_1)(P(k_2)+P(k_3))
\nonumber \\&=& 4\frac{3}{5}\fnl P(k_1)P(k)\left[ 1 + \left(\frac{k_1}{k}\right)^2\frac{9+15\cos(2\phi)}{16} +\dots\right],
\label{b_expand}
\eea
This demonstrates that, in the squeezed limit, the local modulation gives an isotropic bispectrum as expected. The leading angular dependence is quadrupolar (with $m=2$) and only enters at ${\cal O}((k_1/k)^2)$, corresponding to anisotropy generated by large-scale gradients in the modulating field: to leading order all the small-scale modes have their amplitude modulated the same way, independent for their orientation. This remains true for smooth nearly scale-invariant spectra, in general Eq.~\eqref{b_expand} just picks up contributions to the ${\cal O}((k_1/k)^2)$ term that depend on the derivative of the power spectrum.

Local modulations appear in large-scale structure, where the modulating field is a large-scale density mode. Above linear order this will affect how small-scale perturbations grow, and hence modulate their amplitude. However the matter power spectrum on scales larger than the horizon size at matter radiation equality (the `turnover') has rapidly decreasing power, so for the density field this effect is only important for squeezed triangles where all scales are smaller than the turnover (in the $k_1\rightarrow 0$ limit the density bispectrum is zero, see e.g. Ref.~\cite{Desjacques:2010nn}). On the other hand if there is primordial non-Gaussianity in the form of a local modulation by a gravitational \emph{potential}, this will also modulate the amplitude of the small-scale power, and hence modulate the number of objects that reach the critical density to form a galaxy. Since the potential is scale invariant on large-scales (proportional to the comoving density over $k^2$ by the Poisson equation), the large-scale potential modulation of the small-scale density leads to an observable large-scale modulation in the number density of galaxies that form. Exactly how this shows up in the power spectrum and higher-point functions is a little subtle, see Refs.~\cite{Dalal:2007cu,Desjacques:2010nn,Baldauf:2011bh} for details. The simplest effect is actually in the galaxy power spectrum: local primordial non-Gaussianity leads to a scale-dependent bias on large-scales.

Local modulations are also potentially important in the early universe: once large-scale fluctuations are well outside the horizon they can give a somewhat different local evolution in different Hubble patches, which can locally modulate what happens to the smaller-scale modes.

\subsubsection{Squeezed bispectrum from inflation}

For squeezed bispectra, the small-scale modes leave the horizon during inflation significantly later than the large-scale modes.
The large-scale modes therefore effectively modify the background seen by the small-scale modes as they leave the horizon and beyond. In single-field inflation there is a one-to-one mapping between the background Hubble parameter and field values, so the conditions on the surface where a small-scale mode leaves the horizon at $k_{\text{phys}} = H$ are locally identical to what they would be without the large-scale mode~\cite{Maldacena:2002vr,Creminelli:2011rh}. However when we observe angular scales on the sky, we can look at many different Hubble patches; since these patches all have slightly different large-scale field values they have undergone slightly different expansion histories: there is a local perturbation to the scale factor in each patch.

Is this observable even though the physics at horizon exit at all points is the same? It might be helpful to consider the analogy of the CMB fluctuations: if we approximate recombination as sharp so that the universe goes immediately from opaque to transparent, the CMB photons we observe are coming from an equal temperature surface (given by the critical temperature at which electrons and protons recombine). However we do not see the same temperature! On small scales this is because in locations where there is an overdensity --- which is hotter --- the universe locally has to expand for longer to reach the recombination temperature. The photon we observe has therefore seen less expansion (and hence redshifting) between recombination and us, and hence appears hotter. In this picture it is the local scale-factor perturbations at recombination that give rise to the small-scale observed temperature anisotropy, even though the surface itself is equal temperature. In the case of the inflation fluctuations the picture is similar: the large-scale modes affect the amount of expansion between horizon exit and observation, and hence the same physical scale has been stretched by slightly different amounts in different places. The difference in size between the many different Hubble patches is in principle observable; an observed angular scale in different directions corresponds to a different physical size exiting the horizon, and hence a slightly different power spectrum amplitude.

%There are several different ways to interpret the effect of these large-scale modes. Locally it looks like a small shift in time coordinate, which is not directly observable. However what we observe is angular scales on the sky, which can span different Hubble patches in the early universe. If we map a given structure that we see on the sky back to when its small-scale modes left the horizon, the slightly different large-scale modes in different directions leads to a slightly different histories, and hence a slightly different value of the inflaton potential when the small-scale modes left the horizon. If the
%inflaton potential is not exactly flat as the modes leave the horizon, the modes therefore see slightly different Hubble parameters, and hence have slightly different fluctuation amplitudes.  Since the inflation potential must be nearly flat, giving small spectrum index $n_s-1\sim 0 $, the response to shifts in horizon-exit are suppressed by a factor of $n_s-1$, so $\fnl \sim {\cal O}(n_s-1)$~\cite{Maldacena:2002vr}. Since this is very small compared to what can easily be observed, detection of significant primordial squeezed bispectrum could rule out all single-field inflation models. In more complicated, e.g. multi-field models, the evolution of the different patches can be modulated more dramatically, leading to observable signals.

However the effect is small, most obviously because the fluctuation amplitude is ${\cal O}(10^{-5})$, so the modulation is not expected to be at larger than ${\cal O}(10^{-5})$, at least if the inflaton potential is smooth.  The change in power with scale is also small because the spectrum is expected to be nearly scale-invariant: from Eq.~\eqref{squeezed_nonpert}, for a power law spectrum $k^3 P(k) \propto k^{n_s-1}$ we have the well-known result for the curvature perturbation bispectrum in the squeezed limit~\cite{Maldacena:2002vr}  (see the Appendix for derivation)
\be
b(k_1,k_2,k_3)= b_{k_1 k}^m  \approx - \delta_{m0}(n_s-1) P(k_1)  P(k).
\label{maldacena}
\ee
For $n_s\sim 1$ this corresponds to $|\fnl| \ll 1$, so a detection of primordial squeezed non-Gaussianity at a level higher that this would rule out essentially all single-field inflation models. However in models with multiple fields the super-horizon evolution can be modulated in more significant ways by auxiliary fields, and larger non-Gaussian signals are possible (for a review see Ref.~\cite{Byrnes:2010em}).

\section{Trispectrum}

\begin{figure}
\includegraphics[width=8cm]{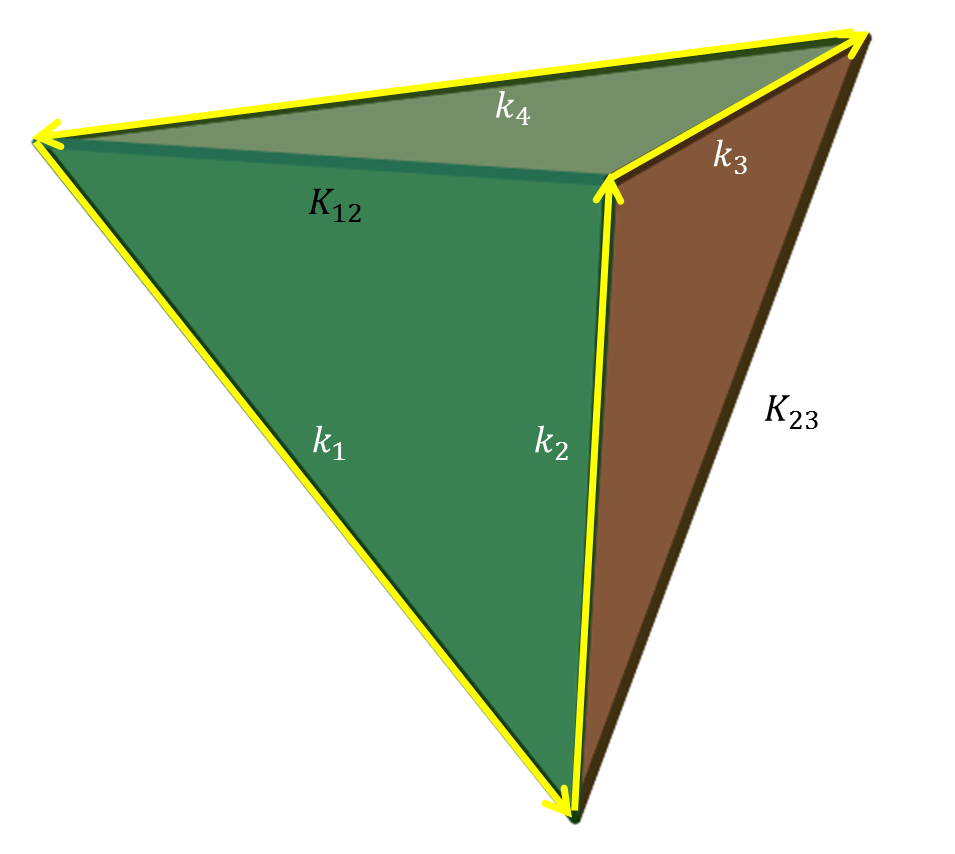}
\caption{A trispectrum shape is defined by four wavenumbers $k_1..k_4$, and the `diagonals' $K_{12}, K_{23}$. Lines with these lengths form the edges of a tetrahedron.
\label{Tetrahedron}
}
\end{figure}

\begin{figure}
\includegraphics[width=15cm]{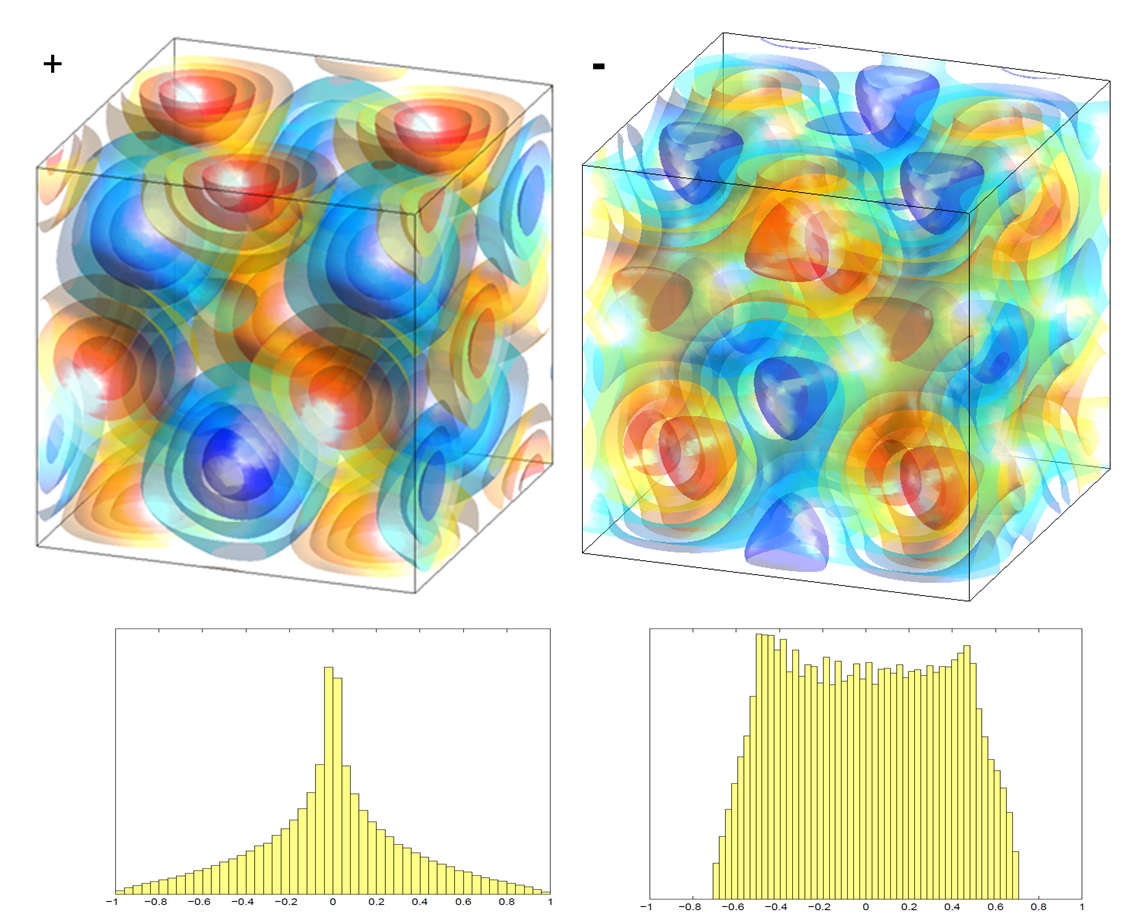}
\caption{Density contours of a three-dimensional field with `tetrahedral' trispectrum, corresponding to the sum of four modes of equal wavenumber, with wavevectors determined by the directions of the vertices of a regular tetrahedron. The figures on the left are for positive trispectrum (positive kurtosis as shown in the histogram of point values); on the right for negative trispectrum (with negative kurtosis). A positive tetrahedral density trispectrum corresponds to having localized regions of concentrated overdensity and concentrated underdensity surrounded by larger volumes that are only weakly perturbed.
\label{Tetrahedral}
}
\end{figure}

The trispectrum is the harmonic four-point function. As with the bispectrum, statistical isotropy and homogeneity means that the four wavevectors must sum to zero, and for a parity invariant ensemble the shape is fully determined by the lengths of six lines joining the four points.
Gaussian fields have a trivial four-point function given by products of the power spectrum, so the trispectrum $\Trispec$ is defined as the more interesting connected part:
\be
\la T(\vk_1) T(\vk_2)T(\vk_3) T(\vk_4)\ra_C \propto \delta(\vk_1+\vk_2+\vk_3+\vk_4) \Trispec(k_1,k_2,k_3,k_4; K_
{12},K_{23}).
\ee
Here $k_1..k_4$ are the lengths of the wavevectors, and $K_{12},K_{23}$ are the lengths of the two `diagonals'; see Fig.~\ref{Tetrahedron}.
%The trispectrum is unchanged under $T\rightarrow -T$, and hence only cannot describe skew (which is described by the bispectrum).
Note than in 3D the points do not all have to lie on a plane: in general the 6 connecting lines define the edges of tetrahedron.

\subsection{Tetrahedral trispectrum}

The most symmetric case is when the tetrahedron defined by the connecting wavenumbers is regular; a field with this kind of `tetrahedral' trispectrum non-Gaussianity is shown in Fig.~\ref{Tetrahedral}. The positive sign corresponds to having concentrated regions of overdensity and underdensity, with weaker perturbations in between. The histograms show the point kurtosis, with sign depending on the sign of the trispectrum. Such a trispectrum could be generated by local dynamics, with strongly non-linear local processes potentially giving a large signal, or inflation models with low sound speed. But note that because the trispectrum is invariant under $T\rightarrow -T$ it does not describe skew, and hence on its own does not describe the effects of gravitational collapse (which generates a large equilateral bispectrum). Of course the tetrahedron does not need to be regular, and non-linear dynamics will generally also generate a tetrahedral signal with similar but not equal sides, corresponding to more stretched-out regions of overdensity and underdensity.

\begin{figure}
\includegraphics[width=13cm]{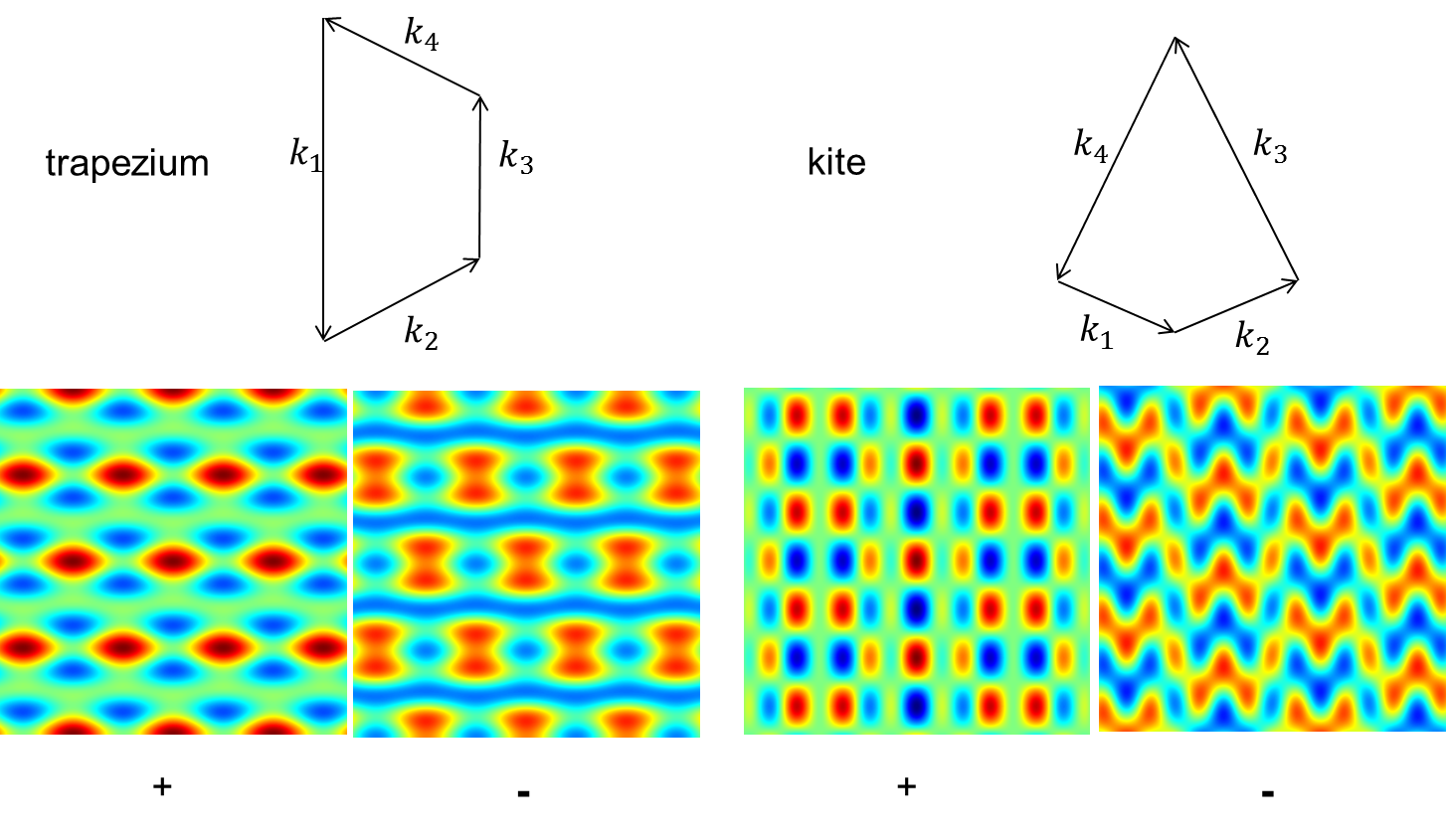}
\caption{Two dimensional fields with trapezoidal and kite-like trispectra of both possible signs. The specific shape trapezium shape has $k_2=k_3=k_3=k_1/2$, and the kite is a cyclic quadrilateral with $2k_1=2k_2=k_3=k_4$.
\label{Quadrilaterals}
}
\end{figure}

In 2D (e.g. on CMB sky) the tetrahedron becomes a quadrilateral. Specific shapes of interest include kites and trapeziums\footnote{Note that I use `trapezium' in the international (non-US) English sense: a quadrilateral with two parallel sides, with adjectival form `trapezoidal'.}~\cite{Hindmarsh:2009es}, symmetric examples of which are shown in Fig.~\ref{Quadrilaterals}. Exact parallelograms are useful for calculating the non-Gaussian contribution to the covariance of the power spectrum, but near-parallelograms are equivalent to the squeezed trispectrum that I discuss in more detail below.

\subsection{Squeezed-diagonal trispectrum}
\label{sec:DiagonalSqueezed}

\begin{figure}
\includegraphics[width=13cm]{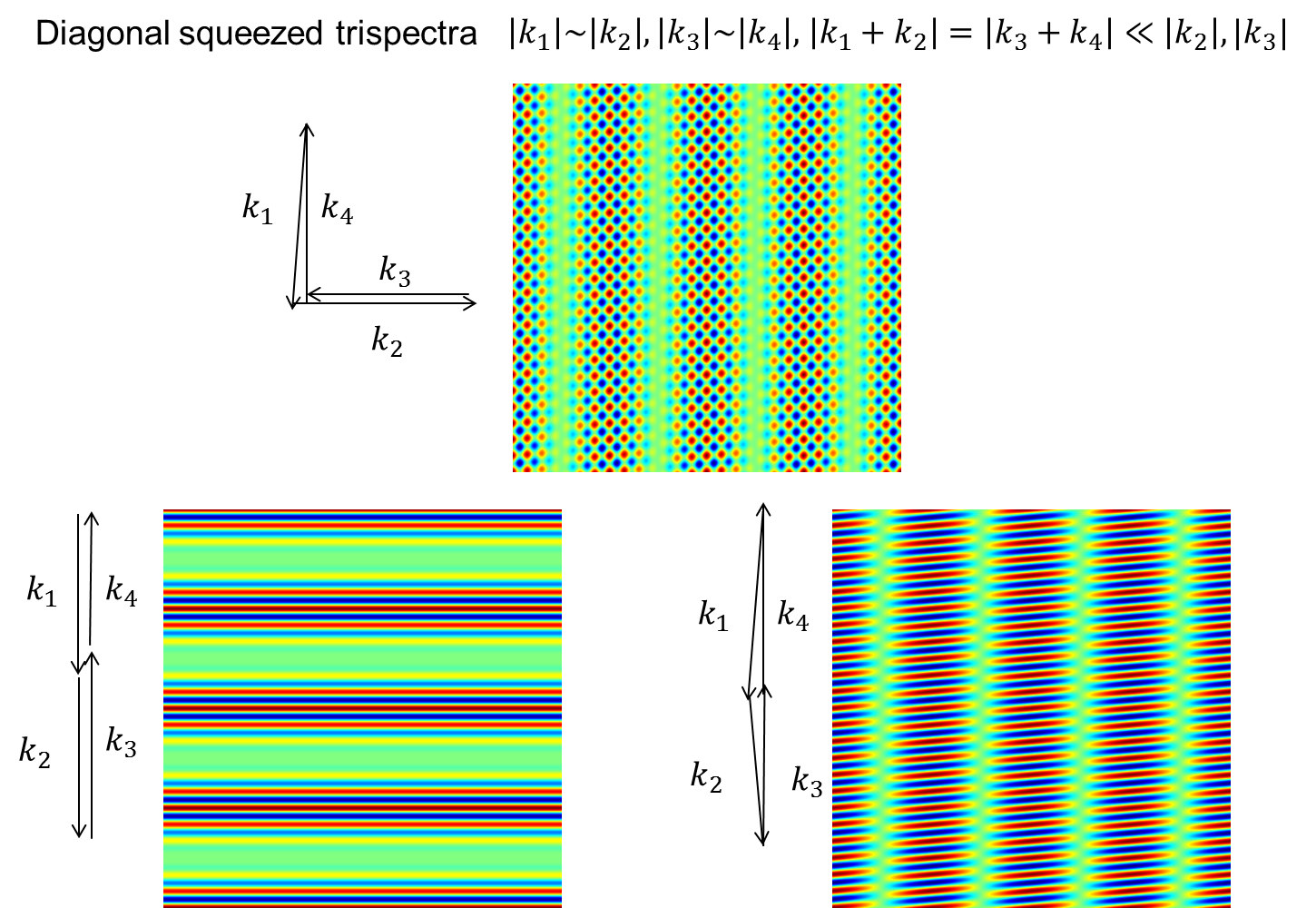}
\caption{Positive trispectra with one short diagonal (diagonal squeezed), corresponding to a large-scale modulation of small-scale power. In 3D fields the wave vectors can also go into the page
\label{DiagonalSqueezed}
}
\end{figure}

As in the case of the bispectrum, the squeezed limit is of particular interest. The diagonal-squeezed trispectrum corresponds to one of the diagonals of the tetrahedron being much smaller than the other sides, e.g. $K \ll k_1,k_2,k_3,k_4$. The tetrahedron can be considered as two triangles having side $K$ being stuck together; hence large-scale wave number $K$ is the wavenumber of the beats seen in Fig.~\ref{TwoWaves}, and corresponds to the wavelength of a large-scale modulation in the small scale power. This is illustrated in Fig.~\ref{DiagonalSqueezed} for wavevectors lying in the same plane; the shape of the tetrahedron determines the relative alignment of the modulation and the small-scale modes. In general the squeezed shape depends on both the diagonals, but in the two-dimensional case, where all the wavevectors lie in a plane, the squeezed-diagonal trispectrum only depends on the short diagonal $K$ since the other diagonal is then determined from the lengths of the sides. Note that an arrangement of wavevectors into an approximately planar parallelogram is equivalent to this configuration, since the wavevectors can be re-ordered so that the nearly parallel vectors are consecutive.

I'll discuss the general form of the theoretical prediction for a squeezed-diagonal trispectrum later in Sec.~\ref{aniso}.
The local modulation of Eq.~\eqref{modulation} can produce it, with all small-scale modes being modulated in the same way. Such a trispectrum is positive (because the modulation field always has positive power spectrum) and isotropic: there is no dependence on the relative alignment of the modes, and the signal in all the configurations shown in Fig.~\ref{DiagonalSqueezed} should be the same. In general however there can be angular dependence; for example in the 2D case of the CMB (where all wavevectors must lie in a plane), CMB lensing shear generates an anisotropic squeezed trispectrum~\cite{Hu:2001fa}. Similarly large-scale gravitational waves produce small orientation-dependent distortions in the small-scale modes~\cite{Seery:2008ax,Masui:2010cz}.

Note that the modulation field here is only observed via its effect on the small-scale modes; as such only the amplitude of the modulation matters, not its correlation to the large-scale field. Further note that if there is a local modulation giving the bispectrum, it must also give a squeezed trispectrum: if there's a spatial modulation of small-scale power correlated to the large-scale temperature giving a bispectrum, the power of the in large-scale modulation field gives a trispectrum. The correlation of the modulation can be between $-1$ and $1$, which limits the size of the bispectrum compared to the trispectrum, with the bispectrum signal being maximized when the modulation field is totally correlated to the large-scale field. Conventionally the size of any primordial isotropic squeezed-diagonal trispectrum is parameterized by a parameter $\taunl$ and there is then an inequality $\taunl \ge (6\fnl/5)^2$ in the squeezed limit~\cite{Suyama:2007bg}. If the amplitude of the modulation depends on $K$, there is also an equivalent relation between the bispectrum with $k_1=K$ and the trispectrum at each $K$, which just says that the correlation of a modulation at any scale with the equivalent large-scale field mode must be less than one.

The diagonal-squeezed trispectrum is hard to measure. Firstly, because if it is due to an $\clo(\epsilon)$ modulation, it is proportional to the power in the modulation which is a tiny $\clo(\epsilon^2)$, so a larger modulation is required to have an observable signal. Secondly because of cosmic variance: the finite number of small-scale modes available with finite resolution observations. If we imagine that the field consists only of large-scale and small-scale modes, to measure the modulation of the small-scale power we need to divide up the space and measure the small-scale power as a function of position. Each box in space will have a sample variance error $\propto 1/n$ in the small-scale power, where $n$ is the number density of the small-scale modes. For boxes large compared to the small-scale modes this error is independent between boxes, and the sample variance therefore looks like white noise. This translates into a white cosmic variance noise on the estimate of the modulation field. If the modulation field is nearly scale-invariant, the sample variance noise therefore dominates on small-scales. In that case only the very large-scale modes of the modulation field have any chance to be measured, and in fact the signal to noise (for small signals) is dominated by the very largest modulations (in the CMB, almost all the signal is in modulations with $L \alt 10$~\cite{Kogo:2006kh}). By contrast the squeezed bispectrum is the correlation of the modulation with the large-scale field, where the large-scale field is typically measured very accurately; so strongly correlated modulations are generally much more easily seen in the bispectrum than trispectrum. However in cases like CMB lensing, where the correlation drops rapidly with scale and the convergence power spectrum is very blue, the trispectrum is easier to detect. Note that if the modulation field starts to be well measured, increasing the number of small-scale modes to lower the sample variance noise does not help any more: the accuracy of the bispectrum and trispectrum measurement is then limited by the cosmic variance of the large-scale modulation modes~\cite{Creminelli:2006gc,Lewis:2011fk}.

\subsection{One-leg and two-leg squeezed trispectra}

\begin{figure}
\includegraphics[width=13cm]{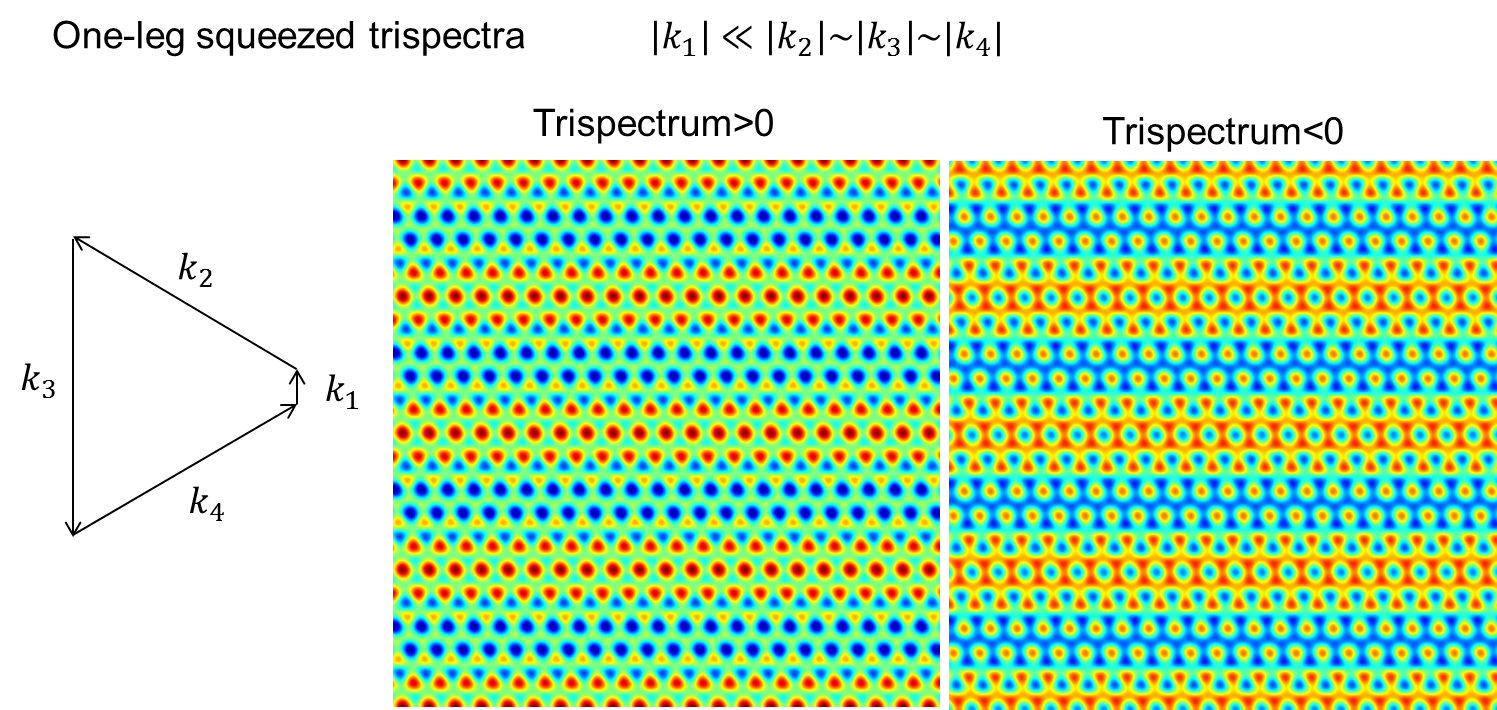}
\caption{Trispectrum with one short wavenumber, $k_1 \ll k_2,k_3,k_4$, corresponding to a large-scale modulation of a small-scale nearly-equilateral bispectrum. In 3D fields the modulation can also go out of the plane so the sign of the density changes along each filament.
\label{OneLegSqueezed}
}
\end{figure}

For the trispectrum there is another squeezed limit to consider: when one of the edges in the tetrahedron is much shorter than the others. A field with this kind of trispectrum is shown in Fig.~\ref{OneLegSqueezed}. The large-scale mode with $k_1\ll k_2,k_3,k_4$ modulates the small-scale modes, and if $k_2\sim k_3\sim k_4$ this look like a modulation of a nearly equilateral bispectrum. As shown in Fig.~\ref{OneLegSqueezed} the small-scales look like concentrated overdensities (positive equilateral bispectrum) or concentrated underdensities (negative equilateral bispectrum), where the sign of the small-scale bispectrum is correlated to the large-scale modulating field value. A positive one-leg squeezed trispectrum corresponds to regions with small-scale concentrated overdensities being more likely where there is a large-scale overdensity, and concentrated underdensities being more likely where there is a large-scale underdensity. For negative trispectrum the association is the other way round.

If the bispectrum that is being modulated is local rather than equilateral, the signal will peak when $k_2\ll k_3,k_4$ and hence in a two-leg squeezed trispectrum shape with $k_1, k_2 \ll  k_3,k_4$. A trispectrum of positive sign can be obtained from a correlated linear local modulation as in Eq.~\eqref{modulation}, and hence is produced by local primordial non-Gaussianity (in addition to a diagonal-squeezed trispectrum). However the negative sign cannot be produced in this way. However a local quadratic modulation can give either sign, for example
\be
\chi = \chi_0(1+\gnl \chi_0^2)
\ee
where $\chi_0$ is Gaussian and $\gnl$ quantifies the size and sign of the modulation. This form of modulation does not produce a bispectrum, but can produce the two-leg squeezed trispectrum of either sign, which can be verified by explicitly calculating terms in the four-point correlation. A negative two-leg squeezed signal is therefore clear signature of this kind of primordial non-Gaussianity. Unfortunately it is rather hard to measure: as we've seen, the signal is a correlation of a bispectrum modulation, so estimating the amplitude effectively requires estimating the bispectrum in different local patches and then correlating its spatial dependence with the large-scale field. The cosmic-variance noise is therefore much larger than when estimating a constant primordial equilateral bispectrum and current constraints are rather weak~\cite{Fergusson:2010gn}.

The general form of the theory one-leg squeezed and two-leg squeezed trispectra follows an analogous argument to that leading to Eq.~\eqref{squeezed_nonpert}: in the linear one short-leg approximation:
\be
\la \tT(\vk_1) \tT(\vk_2) \tT(\vk_3)\tT(\vk_4) \ra \approx \la T(\vk_1) \tT(\vk_2) \tT(\vk_3) \tT(\vk_3)  \ra =  P_{TX_i}(k_1) \left\la \frac{\delta}{\delta X_i(\vk_1)^*} \left(\tT(\vk_2) \tT(\vk_3)\tT(\vk_4)\right) \right\ra.
\label{oneleg_nonpert}
\ee
 The single-field inflationary prediction is very small because this trispectrum depends on changes in the bispectrum, and the bispectrum itself is very small~\cite{Seery:2006vu}. The signal in large-scale structure is expected to be much larger and of positive sign, since non-linear growth gives an equilateral bispectrum, and the amount of non-linear growth will be modulated by large-scale densities. The two-leg squeezed trispectrum can be calculated similarly assuming linearity in the two short-legs, and can be useful for example when calculating analytic approximations for the $\gnl$ trispectrum expected in the CMB~\cite{Pearson:2012ba}.

\subsection{Flattened quadrilateral trispectrum}
\label{strings}

\begin{figure}
\includegraphics[width=12cm]{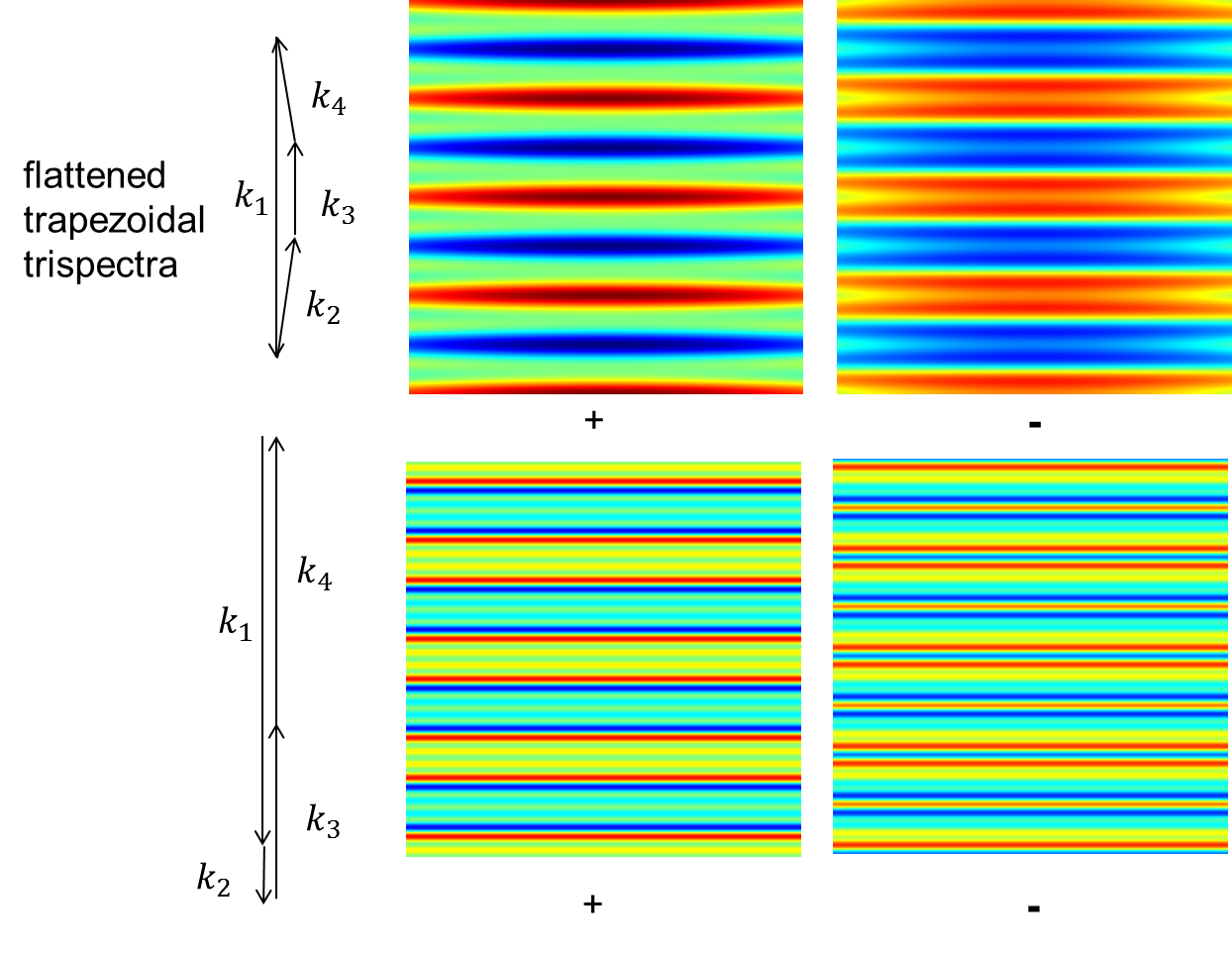}
\caption{Examples of two dimensional fields with flattened trapezoidal trispectrum. In the top figure the waves have equal amplitude, and in the flat limit where $k_2=k_3=k_4=k_1/3$ the negative sign trispectrum would come from the first two terms in the harmonic expansion of a square wave. The lower figure has wavenumbers in the ratio $1,3,5,7$ with equal amplitude; if the amplitudes were in the ratio $1,1/3,1/5,1/7$ the positive trispectrum would come from the first four terms in the harmonic square-wave expansion. Cosmic strings are expected to produce the top trispectrum with negative sign, and the bottom one with positive sign (in agreement with Ref.~\cite{Hindmarsh:2009es}), corresponding to having sharp real-space line-like changes in temperature.
\label{FlattenedTrapezium}
}
\end{figure}

We have already discussed trispectra with a squeezed diagonal, which if all the wavevectors are nearly aligned is an example of a flattened quadrilateral (e.g. the lower fields in Fig.~\ref{DiagonalSqueezed}). In 2D the flattened limit corresponds to having a real-space symmetry along lines. Other cases of interest are the flattened trapezoidal shapes shown in Fig.~\ref{FlattenedTrapezium}.

If there are cosmic strings, the Kaiser-Stebbins effect~\cite{Kaiser:1984iv} can produce line-like discontinuities in the CMB temperature. This is because the strings are very thin, but moving fast: in their rest frame they see a large CMB dipole, which is lensed by the string stress-energy, giving a change in observed temperature across the string. So strings produce sharp lines in the CMB where the temperature suddenly changes from hotter to colder. There is therefore power on very small scales (big change in temperature across the string), with little power in between strings; this is like a very localized modulation of small-scale power, corresponding to a squeezed-diagonal trispectrum shape.

An array of strings with appropriate velocities would give a signal in the CMB that looks like a plane square wave, with the line-like discontinuities corresponding to the edges of the steps. Expanded into Fourier modes, a square wave with wavevector in the $\vx$ direction gives
\be
\sum_n (-1)^n \frac{\cos\left([2n+1] kx \right)}{2n+1} = \cos(k x) - \frac{1}{3}\cos(3k x)+\frac{1}{5}\cos(5 k x)-\frac{1}{7}\cos(7 k x)\dots ,
\ee
 so there is a relation between the amplitudes seen in the modes with wavenumber $k$, $3k$, $5k$ and $7k$. For example $T(k)=-3T(3k)$, and thus we expect $\la T(k) T(3k)T(3k)T(3k)\ra < 0$; likewise we expect $\la T(k) T(3k)T(5k)T(7k)\ra > 0$. This can be seen in Fig.~\ref{FlattenedTrapezium}, where only one of the signs in each configuration looks roughly like an array of line temperature discontinuities as expected from cosmic strings. So cosmic string networks will produce negative trispectra of some flattened shapes, along with trispectra in other flattened (and less-flattened) configurations with positive sign~\cite{Hindmarsh:2009es,Regan:2009hv}. Note that cosmic strings produce a small bispectrum (but significant trispectrum) since the velocities of long strings producing the Kaiser-Stebbins signal can be in any direction, so that to a first approximation $T(\vx)$ and $-T(\vx)$ are equally likely.

Other flattened quadrilateral shapes will of course be produced by other 1D density functions. For example delta-function line sources contribute equally to all wavenumbers in the preferred direction, so there would be a positive flattened trispectrum with wavenumbers in the ratio 1:2:3:4 where $\vk_4+\vk_1=-\vk_2-\vk_3$.

\section{Statistical anisotropy and modulation reconstruction}
\label{aniso}

We've seen that large-scale modulating fields can give rise to a squeezed bispectrum and trispectrum when we average over all possible field values. However we could ask a slightly different question: in our universe there must be a particular realization of the large-scale modulation field, can we learn what it is? Looking at Figs.~\ref{SqueezedAligned} and \ref{DiagonalSqueezed} how to do this seems obvious: the small-scale power is modulated, so if we estimate the power spectrum in different places, the spatial dependence of the power spectrum should trace out the modulation.

Imagine that there are modulation scalar modes $X$ that we want to learn about. If they are large-scale modes, the analog of the linear short-leg approximation here is that products of small-scale fields only depend linearly on $X$ (but possibly non-linearly on all the other fields\footnote{Note that the wavenumber dependence of $X$ can be restricted as much as we like, so this only requires linearity in a single mode: on the CMB the dependence can be non-linear in all modes as long as it is linear in the single large-scale $l,m$ mode being considered.}). For a pair of small-scale modes we therefore have
\be
\tT(\vk_2) \tT(\vk_3) \approx \tT(\vk_2) \tT(\vk_3) |_{X=0} + \int \ud \vK X(\vK)^* \frac{\delta}{\delta X(\vK)^*} \left( \tT(\vk_2) \tT(\vk_3)  \right).
\ee
If we then average over the conditional distribution for the fields given fixed $X$, for $\vk_2\ne -\vk_3$ we have
\be
\la \tT(\vk_2) \tT(\vk_3)\ra_{P(\tT|X)} \approx
\int \ud \vK  X(\vK)^* \left\la \frac{\delta}{\delta X(\vK)^*} \left( \tT(\vk_2) \tT(\vk_3)  \right)\right\ra,
\ee
where for a result to linear order in $X(\vK)$ the expectation on the right hand side can be evaluated averaging over all the fields. If we correlate this with $T(\vk_1)$ and average over $X$ this recovers the general form of the squeezed bispectrum given in Eq.~\eqref{squeezed_nonpert}. From statistical homogeneity the wavevectors in the expectation value over all fields must form a triangle, so this term is proportional to a delta-function, and we can define
\be
\left\la \frac{\delta}{\delta X(\vK)^*} \left( \tT(\vk_2) \tT(\vk_3)  \right)\right\ra \equiv \cla(K,k_2,k_3)\delta(K+k_2+k_3),
\ee
where $\cla(K,k_2,k_3)$ encodes how the small-scale modes change with the large-scale modulation. Hence averaging over all the other modes recovers something just proportional to the modulation $X(\vK)$~\cite{Hu:2001kj,Lewis:2011fk}:
\be
\la \tT(\vk_2) \tT(\vk_3)\ra_{P(\tT|X)} \approx \cla(K,k_2,k_3) \left.X(\vK)^*\right|_{\vK=-\vk_2-\vk_3}.
\ee
This tells us that by averaging over a quadratic combination of all the observed small-scale modes (which all see the same large-scale modulation realization), we can construct an estimator $\hat{X}$ of the modulation field $X$. This is the more formal statement of the obvious idea that measuring the small-scale power as a function of position should trace out the large-scale modulation. The use of such quadratic estimators has proved useful in CMB studies for reconstructing fields in wide classes of statistically anisotropic models~\cite{Pullen:2007tu,Hanson:2009gu}, and for lensing reconstruction (where the modulation field is the large-scale lensing potential~\cite{Zaldarriaga:1998te,Okamoto03}; for reviews see Refs.~\cite{Lewis:2006fu,Hanson:2009kr}). Of course in observations there are only a finite number of small-scale modes to average over, and this leads to a cosmic variance reconstruction `noise': random fluctuations in an isotropic field will look anisotropic, which is hard to distinguish from a modulation-induced anisotropy with only a small number of observed modes. The CMB lensing signal is large enough to be relatively easily detected (as seen recently by ACT~\cite{Das:2011ak}; see~\cite{Okamoto03} for other forecasts), but primordial signals are much more difficult since only the very largest-scale modulations are not swamped by reconstruction noise (as discussed in Sec.~\ref{sec:DiagonalSqueezed}).

For a fixed modulation the field can look anisotropic and inhomogeneous. However if the modulation is a statistically homogeneous and isotropic field, after averaging over the modulation everything is again statistically homogeneous and isotropic. The bispectrum and trispectrum then respectively quantify the correlation and variance the modulation. Specifically, correlating the modulation reconstruction with $T(\vk_1)$ yields an estimate of the bispectrum~\cite{Komatsu:2003iq,Hanson:2009gu,Munshi:2009ik}:
\ba
\la T(\vk_1) \tT(\vk_2) \tT(\vk_3) \ra &=&\left\la  T(\vk_1) \la \tT(\vk_2) \tT(\vk_3) \ra_{P(\tT|X)} \right\ra_{X,T}  \nonumber \\
&=& \delta(\vk_1+\vk_2+\vk_3) P_{TX}(k_1) \cla(k_1,k_2,k_3) ,
\label{aniso-bispectrum}
\ea
which will be non-zero if the modulation and large-scale fields are correlated, $P_{TX}\ne 0$. If the field for fixed modulation is Gaussian, the power spectrum of the modulation is proportional to the squeezed-diagonal trispectrum since
\ba
\la \tT(\vk_1) \tT(\vk_2) \tT(\vk_3) \tT(\vk_4)\ra &=&
\left\la \la \tT(\vk_1) \tT(\vk_2)\ra_{P(\tT|X)} \la \tT(\vk_3) \tT(\vk_4)\ra_{P(\tT|X)}  +\text{perms} \right\ra_X  \nonumber \\
&\approx&
\delta(\vk_1+\vk_2+\vk_3+\vk_4) P_{XX}(\vK)  \cla(K,k_1,k_2) \cla(K,k_3,k_4)
% \left\la \frac{\delta}{\delta X(\vK)^*} \left( \tT(\vk_1) \tT(\vk_2)  \right)\right\ra
%  \left\la \frac{\delta}{\delta X(\vK)} \left( \tT(\vk_4) \tT(\vk_3)  \right)\right\ra
\label{aniso-trispectrum}
\ea
where in the last line I assumed $K=|\vk_1+\vk_2|=|\vk_3+\vk_4| \ll k_1,k_2,k_3,k_4$ is the short side of the tetrahedron (but $K\ne 0$) and a nearly scale-invariant modulation so that $P_{XX}(K) \gg P_{XX}(|\vk_1+\vk_3|),P_{XX}(|\vk_1+\vk_4|)$.

For squeezed non-Gaussianity where Eqs.~\eqref{aniso-bispectrum},\eqref{aniso-trispectrum} hold, since $P_{XT}^2\le P_{XX}P_{TT}$ there is an inequality between the amplitude of the trispectrum with $k_1,k_2,k_3,k_4,K$ and the amplitude of bispectra with wavenumbers $K,k_1,k_2$ and $K,k_3,k_4$ (and the power spectrum $P_{TT}$): if you have a squeezed bispectrum you inevitably have a non-zero trispectrum. For the local model this gives the $\taunl \ge (6\fnl/5)^2$ inequality discussed in Sec.~\ref{sec:DiagonalSqueezed} (for full derivation in the case of scalar modulations see Ref.~\cite{Smith:2011if}, though an equivalent inequality holds more generally). In terms of optimal estimators for the modulation (summing over all the contributing wavenumbers), this inequality is often obvious; for example for CMB lensing it just states that $(C_l^{T\psi})^2 \le C_l^{TT}C_l^{\psi\psi}$, where $\psi$ is the reconstructed lensing potential and $T$ is the CMB temperature.
There are similar inequalities involving higher-point functions; for example, a one-leg squeezed trispectrum corresponds to a correlated bispectrum modulation, so the bispectrum modulation must have some variance and hence there must be a non-zero 6-point function.

\section{Randomly located features}
\label{features}

If there is some particular fixed feature that can randomly appear on the sky, if the location is unknown, then the location-marginalized model is still statistically isotropic  --- there is no preferred direction or orientation of the feature. As such, the feature (or features) will give a signal in non-Gaussianity searches. For a simple example, say there can be a circularly-symmetric feature on the sky with some radial profile $T(r)$ (for example as might be produced by a texture~\cite{Cruz:2007pe} or inflationary bubble collision~\cite{Aguirre:2009ug,Feeney:2010dd}, in addition to other signals).
If it is centred at location $\vc$ then
\ba
T(\vk) &=& \int \frac{\ud \vx}{2\pi} T(\vx-c) e^{-i\vk\cdot \vx} =
e^{-i\vk\cdot c} \int \frac{r \ud r d\theta}{2\pi}  T(r) e^{-ikr\cos\theta} = e^{-i\vk\cdot c} \int r\ud r T(r) J_0(kr)
\nonumber\\
&=&
e^{-i\vk\cdot c} t(k)
\ea
where $t(k)$ is the Bessel-transform of $T(r)$ defined by the previous line. Then if we integrate out the random location of the centre the $n$-point function is
\ba
\la T(\vk_1)\dots T(\vk_n)\ra \propto \int \ud \vc \prod_i t(k_i)  \exp(-i \vk_i \cdot \vc) \propto \delta\left(\sum_i \vk_i\right) \prod_i t(k_i).
\ea
This is a particularly simple separable form of $n$-point function. As a specific simple case, if the feature is a constant-temperature disk of radius $R$, then
\ba
\la T(\vk_1)\dots T(\vk_n)\ra \propto \delta\left(\sum_i \vk_i\right) \prod_i  \frac{J_1(k_i R)}{(k_i/R)}.
\ea
If the template is not rotationally invariant, then of course the orientation also has to be integrated out. If the feature changes sign under a rotation this leads to all of the odd $n$-point functions being zero, as for example in the simplified example of line-like temperature changes due to cosmic strings discussed in Sec.~\ref{strings}. If the amplitude of the feature is also random and symmetric about zero, integrating out the amplitude will also give zero for all the odd $n$-point functions.

If there is a feature with random sign, but there is only one on our sky, on our sky it will have some particular sign, and hence contribute to the bispectrum estimator in that realization. Is this consistent with prediction for the bispectrum being zero? Actually yes, because the variance of the bispectrum estimator is also sensitive to non-Gaussianity, and in this case the apparent signal in the bispectrum estimator can be interpreted as the variance of the bispectrum being inconsistent with the Gaussian expectations (in other words the large bispectrum estimator in our one realization is a detection of the six-point function). In such cases it may be more natural to think about the conditional distribution, where the sky is anisotropic for a given fixed feature location.

Localized features are of course unlikely to be best detected by looking at only the bispectrum and trispectrum, but if they are present then their contributions to these $n$-point functions can be important, and hence must be distinguished from other signals of interest.

\section{Conclusions}

Non-Gaussianity studies can extract significantly more information from cosmological data than just the power spectrum. I've shown what the various types of qualitatively different signal look like in real space, which can be an intuitive aid to understanding the origin of possible non-Gaussian signals and which sources of non-Gaussianity might be confused with each other. I gave some general results for the form of the squeezed bispectrum and trispectrum and the relation between them. Squeezed configurations are especially interesting configuration to study in terms of learning about the physics of inflation. The angular dependence of any the squeezed signal can be used to identify different physical effects. Local modulations as produced by multi-field inflation are isotropic, but anisotropic models, or late-time effects like CMB lensing, can generate a distinctive anisotropic squeezed bispectrum. It can also be useful to think about non-Gaussianities in terms of statistically anisotropy; for a fixed modulation field the universe looks anisotropic, and the large-scale modes of the modulating field can be reconstructed by using observations of many modulated small-scale modes. However reconstructed modulations, and non-Gaussianity estimates generally, are often limited by cosmic variance: the finite number of modes that are available to distinguish random fluctuations in a realization of a purely Gaussian field from genuine non-Gaussianities.  This is usually the limiting factor when trying to detect small primordial signals.

\section{Acknowledgements}
I would like to thank Mustafa Amin and the various other people who encouraged me to make the pretty pictures more widely available. I also thank David Seery and Donough Regan for discussion, and Ruth Pearson and Andrew Liddle for comments and advice on a draft. I acknowledge support from the Science and Technology Facilities Council [grant number ST/I000976/1].
\appendix

\section{Squeezed bispectrum from single-field inflation}

For completeness I show how the well-known result for the single-field squeezed-limit
bispectrum~\cite{Maldacena:2002vr,Creminelli:2004yq,Creminelli:2011rh} can be obtained from the general form the squeezed bispectrum given in Eq.~\eqref{squeezed_nonpert}.
% most simply we can consider the limit $k_1\rightarrow 0$, which obtains
%\be
%\la \tilde{\zeta}(\vk_1) \tilde{\zeta}(\vk_2) \tilde{\zeta}(\vk_3) \ra \approx P_{\zeta\zeta}(k_1)
%\frac{\delta}{\delta \zeta} \left\la \left(\tilde{\zeta}(\vk_2) \tilde{\zeta}^*(\vk_2)\right) \right\ra.
%\ee
In the case of single-field inflation the modulating field is the large-scale curvature perturbation $\zeta(\vk_1)$. If $k_1 \ll k_2,k_3$ the background scalar metric seen by the small-scale modes can be written
\be
ds^2 = dt^2 - e^{2\bar{\zeta}}a(t)^2 d\vx^2 = dt^2 - a(t)^2 d\tilde{\vx}^2
\ee
Here $\bar{\zeta}$ is essentially a local scale-factor perturbation, and in the second step we took the squeezed limit so that $\bar{\zeta}(\vx)=\bar{\zeta}$ is constant over the region of interest, and $\tilde{\vx} = e^{\bar{\zeta}} \vx$ is a re-scaled coordinate. In terms of $\tilde{\vx}$ the metric just looks like a normal background FRW universe, so the non-linear small-scale mode can be obtained from $\tilde{\zeta}(\vx)=\zeta(\tilde{\vx})=\zeta(e^{\bar{\zeta}}\vx)$ where $\zeta(\vx)$ is the usual linear fluctuation obtained in single-field inflation. The response of the small-scale mode to the large-scale perturbation is therefore
\bea
\frac{\delta}{\delta\zeta(\vk_1)^*} \tilde{\zeta}(\vk_2) &=&
 \frac{\delta}{\delta\zeta(\vk_1)^*}\int \frac{\ud\vx}{(2\pi)^{3/2}} \zeta(e^{\bar{\zeta}} \vx) e^{-i \vk_2\cdot \vx}\\
 &=&   \frac{\delta}{\delta\zeta(\vk_1)^*}\int \frac{\ud\vx}{(2\pi)^{3/2}} \int \frac{\ud\vk}{(2\pi)^{3/2}}  \zeta(\vk) e^{i \vk\cdot e^{\bar{\zeta}} \vx} e^{-i \vk_2\cdot \vx}\\
&=&\int \frac{\ud\vx}{(2\pi)^{3/2}} \int \frac{\ud\vk}{(2\pi)^{3/2}} i(\vk\cdot \vx) e^{\bar{\zeta}} \zeta(\vk) \frac{\delta \bar{\zeta}}{\delta\zeta(\vk_1)^*} e^{i \vk\cdot e^{\bar{\zeta}} \vx} e^{-i \vk_2\cdot \vx}
\\
&=& \int \frac{\ud\vx}{(2\pi)^{3/2}} \int \frac{\ud\vk}{(2\pi)^{3/2}} \zeta(\vk)  \frac{\ud}{\ud \ln k}e^{i \vk\cdot e^{\bar{\zeta}}\vx}
\frac{e^{-i\vk_1\cdot \vx}}{(2\pi)^{3/2}} e^{-i \vk_2\cdot \vx}.
\eea
It is possible to proceed non-perturbatively, but for small inflation perturbations it is not necessary, so we can evaluate the response to zeroth order in $\bar{\zeta}$ giving
\bea
\frac{\delta}{\delta\zeta(\vk_1)^*} \tilde{\zeta}(\vk_2) &\approx&
-\int \frac{\ud\vk}{(2\pi)^{3/2}} \delta(\vk-\vk_1-\vk_2) \frac{1}{k^3}\frac{\ud}{\ud \ln k} (k^3\zeta(\vk))\\
&=& \frac{-1}{(2\pi)^{3/2}}\left.\frac{1}{k^3}\frac{\ud}{\ud \ln k} (k^3\zeta)\right|_{\vk=\vk_1+\vk_2}.
\eea
We then have
\be
\la \tilde{\zeta}(\vk_1) \tilde{\zeta}(\vk_2) \tilde{\zeta}(\vk_3) \ra \approx
-\frac{1}{(2\pi)^{3/2}}\delta(\vk_1+\vk_2+\vk_3)
P_{\zeta\zeta}(k_1)\frac{1}{2}\left[ \frac{1}{k_3^3}\frac{\ud}{\ud \ln k_3} (k_3^3P_{\zeta\zeta}(k_3))
 + \frac{1}{k_2^3}\frac{\ud}{\ud \ln k_2} (k_2^3P_{\zeta\zeta}(k_2)) \right].
\ee
For a power law spectrum this gives Eq.~\eqref{maldacena} in the main text. Ref.~\cite{Creminelli:2011rh} gives a more careful treatment of leading gradients in the large-scale field.

%\bibliography{../../antony,../../cosmomc}
\providecommand{\aj}{Astron. J. }\providecommand{\apj}{Astrophys. J.
  }\providecommand{\apjl}{Astrophys. J.
  }\providecommand{\mnras}{MNRAS}\providecommand{\aap}{Astron. Astrophys.}

\end{document}